# Extracting Composition-Dependent Diffusion Coefficients Over a Very Large Composition Range in NiCoFeCrMn High Entropy Alloy Following Strategic Design of Diffusion Couples and Physics Informed Neural Network Numerical Method

Suman Sadhu[1], Saswata Bhattacharyya[2] and Aloke Paul[1]
[1]Department of Materials Engineering, Indian Institute of Science, Bengaluru 560012, India
[2]Department of Materials Science and Metallurgical Engineering, Indian Institute of Technology, Hyderabad, Sangareddy 502284, Telangana, India
*Corresponding authors: Suman Sadhu (sumansadhu@iisc.ac.in), Aloke Paul (aloke@iisc.ac.in), Saswata Bhattacharyya (saswata@msme.iith.ac.in)

## Abstract

Estimating composition-dependent diffusion coefficients in multicomponent alloys was a longstanding challenge due to limitations in experimental methods. In this study, we have first demonstrated a strategic design of producing only three diffusion couples to estimate all types, i.e. tracer, intrinsic, and interdiffusion coefficients at the Kirkendall marker planes. This establishes a systematic variation of diffusion coefficients with composition in a very wide composition range of the NiCoFeCrMn system in comparison to the data available on impurity diffusion coefficients in pure elements and tracer diffusion coefficients at the equiatomic composition estimated by the radiotracer method. The importance of estimating the intrinsic diffusion coefficients for discussing diffusional interactions is elaborated. The role of the vacancy wind effect is explained, showing the importance of considering Onsager's cross-phenomenological constants, which are mostly neglected in experimental and numerical optimization methods. Following, a physics-informed Neural Network (PINN)-based numerical inverse method is developed for the first time in an actual multicomponent diffusion context. Governing physics (Fick's law, Onsager formalism, Boltzmann scaling) is embedded into the loss function, enabling inversion from a single profile. The model is anchored by experimental constraints - measured tracer diffusivities, impurity diffusivities, and boundary conditions, avoiding ill-posed solutions that arise from profile-only fits. This helps to extract composition-dependent diffusivities over the whole composition range of the diffusion couples. The framework is generic, extendable to other HEAs or multicomponent alloy systems, and enables a mesh-free, non-radioactive route to diffusion analysis, circumventing the need for tracer isotopes or intersecting diffusion paths.

## 1. Introduction

Estimating diffusion coefficients following the diffusion couple method in multicomponent systems remains the most challenging task in the field of materials science [1–3]. Until recently, it was textbook knowledge that the diffusion coefficients cannot be estimated in a system of more than three elements. This requires (*n*-1) diffusion paths to be intersected in an *n*-component *i.e.* (*n*-1) dimension space for direct estimation of the interdiffusion coefficients, which is simply impossible because of complicated diffusion paths, which cannot be predicted *a priori* [1-3]. Such complications are not there for a radiotracer method; however, the practice of this method is scarce because of the risk associated with handling radioisotopes and also because of the unavailability of radioisotopes of many important elements with reasonable cost and long enough half-life for conducting the experiment. Therefore, diffusion coefficient measurements are done mostly following the diffusion couple method, and efforts are being made to solve the unsolved problem of estimation of diffusion coefficients in multicomponent systems. Most of the diffusion studies conducted until now are conducted in simpler binary and ternary systems, which do not necessarily represent the diffusion behaviour in multicomponent systems used in applications. Moreover, the interdiffusion coefficients (a kind of average of intrinsic diffusion coefficients but important for direct correlation/calculation of the diffusion profile) mostly estimated in the ternary systems may not be sufficient since we need the tracer diffusion coefficients for the generation of mobility database and intrinsic diffusion coefficients for the understanding of the diffusional interactions of elements. The interdiffusion coefficients, especially in concentrated alloys, do not correctly reflect the diffusional interactions of elements, which is substantiated in this study.

A few efficient methods have recently been proposed to solve the longstanding problem of estimating diffusion coefficients in multicomponent systems following the innovative design of diffusion couples. Morral proposed the body diagonal diffusion couple method for estimating interdiffusion coefficients from (*n*-1) closely passed diffusion paths (even if these do not intersect exactly but pass closely to the body centre composition) [4]. This is possible when the diffusion coefficients of elements are not very different [5, 6]. However, in unpublished research, we found this could be tricky, especially when there is a significant difference in the diffusion coefficient of elements. The diffusion paths may not pass closely enough to the body centre composition, and the distances between the diffusion paths could be significantly high. This will be discussed in another publication in future. The number of diffusion paths (n-1) required to produce with such constraints for estimating interdiffusion coefficients is not feasible with a higher number of elements in a system. Instead of attempting the direct estimation of interdiffusion coefficients from (n-1) diffusion profiles, Dash and Paul [6] extended this method in the line of the Kirkaldy lane method [7] for direct estimation of tracer from interdiffusion fluxes if thermodynamic data are available. This requires only two diffusion paths to intersect or pass closely, which has a benefit over the methods proposed by Morral [4] and Kirkaldy and Lane [7] for first estimating the interdiffusion coefficients, which requires (n-1) diffusion paths to intersect or pass closely. This is not easy in multi-component systems without inducing much higher errors in calculated data since it is expected to increase in error induced by the increase in the number of diffusion profiles needed depending on the number of elements in a system. The intrinsic and interdiffusion coefficients can then be calculated from the tracer diffusion coefficients. One can also reduce the experimental effort significantly in combination with the constraint-enhanced numerical methods [8-10] for extracting composition-dependent diffusion coefficients. Moreover, we have already substantiated that tracer and intrinsic diffusion coefficients are important to estimate for understanding the relative mobilities, atomic mechanism of diffusion and diffusional interactions between elements, especially in concentrated or high entropy alloys [11]. The interdiffusion coefficients are important parameters to correlate with the diffusion profiles directly; however, discussing diffusional interactions with these data can be misleading, especially in Multi-principal Element Alloys (MPE) or concentrated alloys [10-14]. These diffusion coefficients are vague in this respect (for discussing the atomic mechanism of diffusion or diffusional interactions between elements) since these are an average of *n* intrinsic diffusion coefficients in an *n*-component system [10-12]. On the other hand, most diffusion studies until recently concentrated on estimating interdiffusion coefficients only, except very few, even for extracting data following a numerical inverse method [15-22]. However, as substantiated in this study, the knowledge of tracer diffusion coefficients is necessary for generating a reliable mobility database.

In the meantime, Paul et al.'s method of producing constrained diffusion profiles (such as pseudo-binary [23-27], pseudo-ternary [24,28], pseudo-quaternary [12] changed the mindset of producing diffusion couples by restricting diffusion paths for the ease of intersecting (which was considered impossible) or passing closely [1-2]. The augmented equation scheme was proposed depending on the types of diffusion profiles used to estimate all the diffusion coefficients from only two diffusion profiles [12-14]. These methods are suitable for estimating diffusion coefficients in any composition range (including concentrated or multi-principal element alloys). On the other hand, a method of estimation from a single diffusion profile is very recently proposed in ternary systems by Xia et al. [29] and Sadhu et al. [11, 30] (independently at almost similar times without knowing each other's work) and in multi-component systems by Sadhu et al. [11]. Sadhu et al. also provided a guideline for producing such diffusion profiles, inducing relatively small errors [11, 30]. This method is especially suitable for alloys rich in a particular element, such as Ni-, Co- or Fe-based, which are currently used in various applications and practised in this study.

The studies on multicomponent diffusion were primarily concentrated in the last decade to test the "sluggish diffusion" hypothesis in high entropy alloys. Initially, it was thought to be proven true [31], although based on faulty analysis [8, 32]. The group of Divinski [33-38] demonstrated that higher entropy does not guarantee the sluggishness of diffusion coefficients with the increase in the number of elements. The difference in diffusion rate is material system specific, which may increase or decrease depending on the type of alloying elements added to the systems [12]. However, this led to several studies developing new methods and analyses of the diffusion couple technique and highlighting in-depth analyses of diffusion mechanisms by solving the unsolved problems of the last several decades [4,6,11-14, 23-30]. With the increase in the number of elements in a system, the composition space of a multicomponent system increases significantly. However, diffusion studies until now have been conducted occasionally in a small range and mostly at or around the equiatomic composition. This is a major limitation of diffusion studies in multicomponent systems. Following the diffusion couple method, we can now estimate diffusion coefficients in ternary and multicomponent systems at the intersecting composition of two [6, 12-14] or ($n$-1) [4, 5] diffusion paths. On the other hand, it can be estimated at one composition of Kirkendall marker plane position [11, 29, 30]. One can estimate composition-dependent diffusion coefficients of all the elements along the diffusion paths following the radiotracer method [39, 40] but on the condition that suitable radioisotopes of all the elements with reasonable price and suitable half-life are available. Moreover, more than one sample must be produced since the radioisotopes' diffusion rate cannot be conducted in a single diffusion couple depending on the difference in types of emitting radiations. As already mentioned, the practice of this method is very limited due to health hazards and limited accessibility (in fact, at present, only one group in the whole world for diffusion studies).

In this article, we have shown a strategic design of diffusion couples such that from only three diffusion couples in the NiCoFeCrMn system, we can estimate and extract diffusion coefficients over the whole composition range of diffusion couples of Ni, Co and Fe content (zero to hundred per cent), Cr content up to 20 at.% and Mn content up to 15 at.% covering the whole composition range of the diffusion couples designed in this study. Diffusion couples are designed for the limited composition of Cr and Mn because of the limited solubility of these elements in the FCC phase of this system. This can be practised for the elements' whole composition range in another suitable system with complete solubilities. This method, leveraging the estimation of all types of diffusion coefficients (intrinsic, tracer, and interdiffusion) from a single diffusion profile at the Kirkendall marker plane, is coupled with a novel Physics-based Neural Network (PINN) optimization developed for the multi-component system in this study. The PINN framework embeds the governing diffusion equation in the training of a deep neural network evaluating the residual of Fick's law at collocation points utilizing experimental constraints (e.g. tracer diffusivities, impurity diffusivities, and boundary conditions) as additional loss terms. The network thus "learns" a function for the composition profile and the diffusivities that simultaneously fit the data and satisfy the physics for the diffusion process. A further cornerstone of our proposed method is the Boltzmann transformation, a classical technique that reduces Fick's second law (a PDE) to an ordinary differential equation (ODE) by exploiting the self-similarity of diffusion profiles over time. Therefore, we demonstrate how (i) the single-couple Kirkendall marker plane method and (ii) a PINN-based inverse model, enhanced by the Boltzmann transformation, can be integrated into a unified framework to generate a composition-dependent mobility database for the whole composition range of the diffusion couples in the NiCoFeCrMn system. Specifically, we show that combining minimal yet strategic experiments with advanced computational modelling yields a full spectrum of tracer, intrinsic, and interdiffusion coefficients from only one diffusion couple. We have shown that the extraction of data by matching the composition (diffusion) profiles only, which is practiced mostly, does not guarantee the extraction of reliable tracer diffusion coefficients. Moreover, the role of the vacancy wind effect, i.e. the contribution of Onsager's cross-phenomenological constants, is described by estimating these parameters, justifying the developing numerical method considering these parameters, which is largely neglected in most of the experimental and computational studies.

## 2. Experimental and Numerical methods for estimation and optimization

### 2.1 Experimental method of producing diffusion couples

Pure elements (99.95 - 99.99 wt.%) were used to produce the $(NiCoFeCr)_{85}Mn_{15}$ alloy in an arc melting unit under an argon atmosphere. Melting was repeated 5-6 times for better remixing of the elements. Following, the alloy buttons were homogenized at 1200ºC for 50 h in a vacuum furnace (~ $10^{-4}$ Pa). After metallographic preparations, spot analysis by WDS (wavelength dispersive spectroscopy) in EPMA (electron probe micro analyzer with W or FEG gun) at random positions was conducted to measure the average compositions. The attempted composition of the alloys is $(NiCoFeCr)_{85}Mn_{15,}$ and the actual alloy composition is measured as $Ni_{20.7}Co_{20.8}Fe_{22.0}Cr_{21.6}Mn_{14.9}$ with an error range of 0.1-0.2 at.%. This is the typical error range of EPMA measurements. Thin slices of ~1.5 mm were cut from the melted button using EDM (electro-discharge machine). After standard metallographic preparation, this was diffusion coupled with pure elements Ni, Co and Fe in a special fixture at 1200ºC for 50 h in a vacuum tube furnace. Before coupling, inert particles, i.e. yttria oxides with an average particle size of 1-2 μm, were applied on one of the end members. Oxide particles were dispersed in acetone, and a small drop was placed on the polished surface. After the evaporation of acetone, the particles were found distributed on the surface. These were then diffusion annealed to produce the interdiffusion zone. After the experiment, the diffusion couples were cross-sectioned by a slow-speed diamond saw. These were then prepared metallographically for WDS line profile measurements in the EPMA. The location of the Kirkendall marker plane was identified by the presence of yttria particles along this plane. Further descriptions of producing diffusion couples with markers can be found in Ref. [2].

**2. 2 Estimation of diffusion coefficients from diffusion couple experiments**

In this article, we have followed the direct estimation method of the tracer diffusion coefficients at the Kirkendall marker plane from a single diffusion profile. The intrinsic fluxes of elements at the Kirkendall marker plane can be calculated by extending the analysis from the binary diffusion couple. [2, 41-43]

$$V_m J_i = -\frac{1}{2t}\left[N_i^+ \int_{x^-}^{x_K} Y_i dx - N_i^- \int_{x_K}^{x^+} (1 - Y_i)\, dx\right], \quad (1)$$

where $J_i$ (*mole/m²s*) is the intrinsic flux of element $i$, $N_i^-$ and $N_i^+$ are the composition of this element in the left and right-hand sides of the diffusion couple, respectively, $t$ is the annealing time, $V_m$ (m³/mol) is the constant molar volume considered, $x_K$ is the location of the Kirkendall marker plane in which distance is measured along $x$ ( in *m*), $x^-$ and $x^+$ are locations at the unaffected left and right-hand side of the diffusion couple, $Y_i = \frac{N_i - N_i^-}{N_i^+ - N_i^-}$ is the Sauer-Freise composition normalized variable [44].

The intrinsic fluxes are related to intrinsic diffusion coefficients in a multicomponent system by

$$V_m J_i = -\sum_{j=1}^{n-1} D_{ij}^n \frac{\partial N_j}{\partial x}, \quad (2)$$

where $D_{ij}^n$ (m²/s) is the intrinsic diffusion coefficient of element *i* related to the composition gradient of element *j*. Therefore, the intrinsic flux of a component is related to ($n$-1) intrinsic diffusion coefficients in which element $n$ is considered as the dependent variable.

The intrinsic diffusion coefficients are related to the tracer diffusion coefficients ($D_i^*$) of elements considering Onsager's cross-phenomenological constants [45,46] following the correlation established by Manning [47].

$$D_{ij}^n = \frac{N_i}{N_j} D_i^* \emptyset_{ij}^n \left(1 + W_{ij}^n\right) \text{ such that } W_{ij}^n = \frac{2}{M_o \emptyset_{ij}^n} \frac{\sum_{i=1}^n N_i D_i^* \emptyset_{ij}^n}{\sum_{i=1}^n N_i D_i^*}, \tag{3}$$

where $\left(1 + W_{ij}^n\right)$ is the vacancy wind effect, a factor that originated from Onsager's cross terms, $M_o$ is a structure factor equal to 7.15 for FCC alloy that is considered in this study [47,48]. and $\emptyset_{ij}^n$ is the thermodynamic factor related to activity by $\emptyset_{ij}^n = \frac{\partial ln a_i}{\partial ln N_1} - \frac{N_1}{N_n} \frac{\partial ln a_i}{\partial ln N_n}$.

We can express the intrinsic flux of an element directly with tracer diffusion coefficients by substituting Eq. 3 in Eq. 1

$$V_m J_i = -\sum_{j=1}^{n-1} \left[\frac{N_i}{N_j} D_i^* \emptyset_{ij}^n \left(1 + W_{ij}^n\right)\right] \frac{\partial N_j}{\partial x} \tag{4}$$

Since we have $n$ intrinsic fluxes related to n tracer diffusion coefficients in a n-component system, we can estimate the tracer diffusion coefficients directly at the Kirkendall marker plane by calculating the intrinsic fluxes for each element. The intrinsic diffusion coefficients ($\widetilde{D}_{ij}^n$) can be then calculated from Eq. 3, and the interdiffusion coefficients can be calculated from [2, 49]

$$\widetilde{D}_{ij}^n = D_{ij}^n - N_i \sum_{k=1}^n D_{kj}^n \tag{5}$$

The advantage of this estimation method and sequence of different types of diffusion coefficients can be understood as one can estimate all types of diffusivities from a single diffusion profile. Estimating interdiffusion coefficients would need (n-1) diffusion paths to intersect or pass very closely, which is difficult.

### 2.3 PINN-based numerical method for optimization

We employ a physics-informed neural network (PINN) framework, implemented via the open-source scientific machine learning library DeepXDE [50], to solve the Boltzmann-transformed diffusion equations under various physical constraints. Unlike the closed-source Parameter Optimization tools available commercially, this open-source approach offers greater flexibility: users can select custom forms for composition-dependent diffusivities, incorporate constraints (e.g., tracer diffusion at specific compositions), and adjust network hyperparameters for improved accuracy.

*2.3.1 Composition-Dependent Diffusivities and Design Parameters*

In a multicomponent alloy system, tracer diffusion coefficients $D_i^*$ depend on the local composition. To represent this dependence compactly and facilitate numerical learning, we model the logarithm of the tracer diffusivity using a polynomial in the independent component fractions. Crucially, we perform all optimization in the non-dimensional form to ensure consistency and numerical stability. Specifically, we define the dimensionless tracer diffusivity as:

$$\overline{D}_i^* = \frac{D_i^*}{D_0} \quad (6a)$$

where ($D_0 = \frac{L_n^2}{T_n}$) in $m^2/s$ is a characteristic diffusivity scale set by a reference length, $L_n$ ($m$) and a reference time $T_n$ ($s$). The relation between the dimensional and nondimensional diffusivities is then:

$$D_i^* = D_0 \cdot \overline{D}_i^* \quad (6b)$$

We assume the following expression for the composition dependence (for *n*-1 compositions as independent variable):

$$\ln(\overline{D}_i^*) = \theta_0^i + \sum_{k=1}^{n-1} \theta_k^{1,i} N_k + \sum_{k=1}^{n-1} \theta_k^{2,i} (N_k)^2 + \sum_{j=1}^{n-2} \sum_{k=j+1}^{n-1} \theta_{k,j}^{3,i} N_k N_j \quad (7a)$$

where $\Theta = \{\theta_0^i, \theta_k^{1,i}, \theta_k^{2,i}, \ldots, \theta_{k,j}^{3,i}\}$ is the set of trainable design parameters, and $N_k$ are the independent component fractions (with one component eliminated using the constraint ($\sum N_i = 1$).

For a five-component system (e.g., NiCoFeCrMn considered in this study), this expands to:

$$\ln(\overline{D}_i^*) = \theta_0^i + \theta_1^{1,i} N_1 + \theta_2^{1,i} N_2 + \theta_3^{1,i} N_3 + \theta_4^{1,i} N_4 + \theta_1^{2,i} (N_1)^2 + \theta_2^{2,i} (N_2)^2 + \theta_3^{2,i} (N_3)^2 + \theta_4^{2,i} (N_4)^2 + \theta_{2,1}^{3,i} N_2 N_1 + \theta_{3,1}^{3,i} N_3 N_1 + \theta_{4,1}^{3,i} N_4 N_1 + \theta_{3,2}^{3,i} N_3 N_2 + \theta_{4,2}^{3,i} N_4 N_2 + \theta_{4,3}^{3,i} N_4 N_3 \quad (7b)$$

This polynomial structure ensures that all inputs and outputs are nondimensional, preserving consistency across the PINN framework. Moreover, since this formulation embeds $\ln(\overline{D}_i^*)$ in the Onsager and flux expressions—where it appears alongside gradients of chemical potentials ($\mu_i$)—each parameter ($\theta$) in Eq. 7a carries units inverse to those of ($RT$) *i.e.* (J/mol)$^{-1}$ so that the exponent remains dimensionless. This treatment aligns with standard thermodynamic modelling practices and avoids inconsistencies that arise when mixing physical units within exponentials.

Finally, this non-dimensional formulation is well-suited to optimization within DeepXDE, offering a robust and interpretable path to learn diffusion coefficients that obey thermodynamic and kinetic constraints. It also provides a natural basis for comparing results across temperature or length scales

*2.3.2 Intrinsic Flux via Onsager Formulation*

For an n-component system, the intrinsic flux $J_i$ of species *i* can be written as [2,45,46, 49]:

$$J_i = -\sum_{k=1}^{n} L_{ik} \frac{\partial \mu_k}{\partial x} \quad (8a)$$

where $\mu_K$ is the chemical potential of the component $k$, and $L_{ik}$ are the Onsager transport coefficients. Note here that $k = i$ represents the main term as well as $k \neq i$ represents the cross terms. Equivalently, one may split off the $k = i$ term to show the main and cross contributions explicitly:

$$J_i = -L_{ii} \frac{\partial \mu_i}{\partial x} - \sum_{\substack{s=1 \\ s \neq i}}^{n} L_{is} \frac{\partial \mu_s}{\partial x} \quad (8b)$$

Manning established the correlations between phenomenological constants and tracer diffusion coefficients expressed as [48].

$$L_{ii} = \frac{C_i D_i^*}{RT} (1 + \xi N_i D_i^*) \qquad L_{is} = \frac{C_s D_s^*}{RT} (\xi N_i D_i^*), \quad (i \neq s) \quad (9)$$

where $\xi = \frac{2}{M_0 \sum_m N_m D_m^*}$ and $M_0$ depends on the crystal structure (7.15 for FCC phase). Substituting Eq. 9 in Eq. 8b, one obtains an intrinsic and tracer diffusion coefficient correlation as:

$$V_m J_i = -\frac{N_i D_i^*}{RT}\frac{\partial \mu_i}{\partial x}(1 + \xi N_i D_i^*) - (\xi N_i D_i^*) \sum_{s \neq i} \frac{N_s D_s^*}{RT}\frac{\partial \mu_s}{\partial x} \tag{10a}$$

or equivalently,

$$V_m J_i = -\frac{N_i D_i^*}{RT}\frac{\partial \mu_i}{\partial x} - \frac{\xi N_i D_i^*}{RT}\left[N_i D_i^* \frac{\partial \mu_i}{\partial x} + \sum_{s \neq i} N_s D_s^* \frac{\partial \mu_s}{\partial x}\right]$$

$$= -\frac{N_i D_i^*}{RT}\frac{\partial \mu_i}{\partial x} - \frac{\xi N_i D_i^*}{RT}\sum_{k=1}^{n} N_k D_k^* \frac{\partial \mu_k}{\partial x} \tag{10b}$$

Furthermore, the interdiffusion flux $\tilde{J}_i$ relates to the intrinsic flux via [2, 49]:

$$\tilde{J}_i = J_i - N_i \sum_{k=1}^{n} J_k \tag{11}$$

Fick's second law for constant molar volume can be expressed as

$$\frac{\partial N_i}{\partial t} = -\frac{\partial (V_m \tilde{J}_i)}{\partial x} \tag{12}$$

Since we analyze diffusion couple after experiment for a certain annealing time, we convert Eq. 12 into an ordinary differential equation (ODE) by scaling with Boltzmann parameter ($\lambda = \frac{x}{\sqrt{t}}$) as

$$-\frac{\lambda}{2}\frac{dN_i}{d\lambda} + \frac{d[V_m \tilde{J}_i(\lambda)]}{d\lambda} = 0 \tag{13}$$

To avoid large numerical values, λ is often normalized to $\bar{\lambda} \in [0,1]$, such that we have

$$-\frac{\bar{\lambda}}{2}\frac{\partial N_i}{\partial \bar{\lambda}} + \eta \frac{\partial[V_m \tilde{J}_i(\bar{\lambda})]}{\partial \bar{\lambda}} = 0, \quad \bar{\lambda} = \frac{\lambda - \lambda^-}{\lambda^+ - \lambda^-}, \quad \eta = \frac{t}{(x^+)^2}, \tag{14}$$

where $\lambda^-$ and $\lambda^+$ are the Boltzmann parameters at the left and right-hand side unaffected part of the end-members of the diffusion couple.

Eq. 14 is the core ODE used in our PINN-based optimization.

*2.2.3 Problem Statement: ODE-Constrained Optimization*

Traditional numerical inverse methods often start from the initial condition and seek to match a transient evolution, yet in practice, experimental composition data are usually measured just one final time at a given temperature. By employing the Boltzmann transformation, we reduce the governing PDE to an ODE in the transformed coordinate $(\bar{\lambda})$, allowing us to directly extract composition-dependent diffusion coefficients from that single end-time snapshot. Consequently, we do not require an incremental time-stepping procedure, making the approach particularly efficient and well-suited to systems where only a final-profile dataset is available.

Let $N_i(\bar{\lambda})$ denote the composition profile of species $i$ in this Boltzmann domain. Our goal is to minimize a composite loss function $\mathcal{L}(\theta, N_i)$ while respecting the following system of constraints:

$$-\frac{\bar{\lambda}}{2}\frac{\partial N_i}{\partial \bar{\lambda}} = \eta\, \nabla \cdot \left[V_m \tilde{J}_i(\bar{\lambda})\right], \qquad \nabla \cdot N_i(\bar{\lambda}) = 0,\ D_i^* = w(D, N_i) \tag{15}$$

Here, $D_i^*$ represents the tracer diffusivity of component $i$, modeled through polynomial expansions given in Eq. 7a in general or Eq. 7b suitable for the quinary alloy assessed in this study. Additional constraints of experimentally estimated data, such as the tracer diffusion coefficients estimated at the Kirkendall marker plane and the impurity diffusion coefficients, are useful/mandatory for reliable optimization compared to optimization based on only the diffusion profiles of elements, which is explained in this study in the results and discussion section. In a five-component system, each $D_i^*$ can be captured by a 15-term polynomial, resulting in up to 75 parameters for all elements combined—though fewer or additional cross-terms may be included at the researcher's discretion.

*2.2.4 Loss Function and Its Components*

We decompose the total loss $\mathcal{L}_{\text{total}}$ into four parts:

$$\mathcal{L}_s = \frac{1}{P_s}\sum_{p=1}^{P_s}\left[-\frac{\bar{\lambda}}{2}\frac{\partial N_i}{\partial \bar{\lambda}} + \eta\,\frac{\partial\left(V_m \tilde{J}_i(\bar{\lambda})\right)}{\partial \bar{\lambda}}\right]^2, \tag{16a}$$

$$\mathcal{L}_b = \frac{1}{P_b}\sum_{p=1}^{P_b}\left|\nabla \cdot \left[N_1(\bar{\lambda}_p)\right] - g_p\right|^2, \tag{16b}$$

$$\mathcal{L}_d = \sum_{p=1}^{P_d}\frac{1}{P_d}\left|\bar{N}_1(\bar{\lambda}_p) - N_1^p(\theta_j)\right|^2, \tag{16c}$$

$$\mathcal{L}_c = \frac{1}{P_c}\sum_{p=1}^{P_c}\left|\bar{w}(\bar{\lambda}_p) - w_p(\Theta)\right|^2, \tag{16d}$$

so that, we have

$$\mathcal{L}_{\text{total}} = \omega_s \mathcal{L}_s + \omega_b \mathcal{L}_b + \omega_d \mathcal{L}_d + \omega_c \mathcal{L}_c \qquad (16e)$$

Here:

- $(\mathcal{L}_s)$ is the PDE residual loss, $(\mathcal{L}_b)$ is the boundary condition loss, $(\mathcal{L}_d)$ is the data fitting loss, and constraint loss $(\mathcal{L}_c)$ is the constraint loss.
- $P_s, P_b, P_d, P_c$ are the collocation points for the PDE/ODE residual, boundary constraints, data points (from experimental composition profiles), and additional constraints, such as tracer and impurity diffusion coefficients, respectively.
- $\bar{N}_1(\bar{\lambda}_p)$ are experimental compositions at location $\bar{\lambda}_p$; $N_1^p(\theta_j)$ are the corresponding model predictions.
- $\bar{w}(\bar{\lambda}_p)$ represents measured tracer or impurity diffusivities at special locations (e.g.\ end members or Kirkendall marker).
- $\omega_s, \omega_b, \omega_d, \omega_c$ are the weights assigned to each loss term, respectively.

The weights are chosen to ensure a robust optimization.

*2.2.5 Manual Loss Weighting for Improved Optimization*

In our inverse modelling framework, the total loss comprises multiple components—PDE residual loss $(\mathcal{L}_s)$, boundary condition loss $(\mathcal{L}_b)$, data fitting loss $(\mathcal{L}_d)$, and constraint loss $(\mathcal{L}_c)$—each associated with a corresponding weight $(\omega_s, \omega_b, \omega_d, \omega_c)$. If these weights are not properly balanced, one loss term (typically the PDE residual) can dominate early training, leading the optimizer to neglect critical experimental constraints. Although DeepXDE offers adaptive re-weighting based on gradient magnitudes, we found this approach to be inefficient and often unstable in our high-dimensional setting.

To address this, we manually tuned the weights across a series of test cases. As shown in Table 1, equal weighting (Case 1) resulted in high final losses. Progressive down-weighting of $(\omega_s)$ and $(\omega_b)$ significantly improved convergence. The best performance was observed in Case 5, with $(\omega_s = 10^{-4}, \omega_b = 10^{-2}, \omega_d = 1, \omega_c = 10^{-5})$, yielding the lowest training and test losses of $4.03 \times 10^{-5}$ and $4.07 \times 10^{-5}$, respectively. These results underscore the critical importance of deliberate, empirically guided weight selection in stabilizing the optimization and achieving physically reliable solutions.

| Cases | $W_s$ | $W_b$ | $W_d$ | $W_c$ | Total training loss | Total test loss |
|---|---|---|---|---|---|---|
| 1 | 1 | 1 | 1 | 1 | $3.59 \times 10^{-1}$ | $3.58 \times 10^{-1}$ |
| 2 | $1 \times 10^{-4}$ | 1 | 1 | 1 | $2.13 \times 10^{-1}$ | $2.13 \times 10^{-1}$ |
| 3 | $1 \times 10^{-4}$ | $1 \times 10^{-2}$ | 1 | 1 | $1.97 \times 10^{-2}$ | $1.97 \times 10^{-2}$ |
| 4 | $1 \times 10^{-4}$ | $1 \times 10^{-2}$ | 1 | $1 \times 10^{-3}$ | $1.86 \times 10^{-3}$ | $1.86 \times 10^{-3}$ |
| 5 | $1 \times 10^{-4}$ | $1 \times 10^{-2}$ | 1 | $1 \times 10^{-5}$ | $4.03 \times 10^{-5}$ | $4.07 \times 10^{-5}$ |

Table 1: Optimizing the weights. Case 5 provides the lowest training and test loss.

*2.2.6 Gradient-Based Adjoint-State Optimization*

To minimize $\mathcal{L}_{\text{total}}$ with respect to the state variables $N_1^s$ and design variables $\Theta$, we utilize a gradient-based method with automatic differentiation (AD). Let $\mathcal{J}(\theta, N_i) = 0$ represent the ODE constraints. We rewrite the composite loss $\mathcal{L}(\theta, N_i(\theta))$ and compute partial derivatives via the adjoint-state technique:

$$\frac{d\hat{\mathcal{L}}(\theta)}{d\theta_j} = \nabla_{\theta_j} \mathcal{L}(\theta, N_i(\theta)) + \nabla_{N_i} \mathcal{L}(\theta, N_i(\theta)) \frac{dN_i(\theta)}{d\theta_j}, \quad j = 1, 2, \ldots. \tag{17a}$$

From the constraint, $\mathcal{J}(\theta, N_i) = 0$ we infer

$$\nabla_{\theta_j} \mathcal{J} + \nabla_{N_i} \mathcal{J} \frac{dN_i(\theta)}{d\theta_j} = 0, \tag{17b}$$

yielding explicit expressions for $\frac{dN_i(\theta)}{d\theta_j}$. Substituting back provides an analytical gradient for the total loss. DeepXDE [50] automates these steps, and it supports both strong and weak forms of the ODE solution.

*2.2.7 Methodology Highlights*

(i) Thermodynamic Integration and Vacancy-Wind Correction

Our method incorporates thermodynamic effects directly into the flux formulation through the chemical potentials $\mu_i$, which naturally capture non-ideal solution behavior and inter-element interactions in concentrated multicomponent alloys. These composition-dependent driving forces are essential for accurately modeling diffusion in such systems. In addition, the Manning relation accounts for vacancy-wind effects, enabling realistic treatment of cross-diffusion by adjusting fluxes when species diffuse at unequal rates. Together, the chemical potential gradients and vacancy-wind corrections form a consistent and physically grounded framework that faithfully captures both thermodynamic and kinetic coupling-critical for simulating diffusion in strongly interacting multicomponent alloys.

*(ii) Robustness to Initial Guesses*

Unlike many inverse modelling approaches that require careful initialization to ensure convergence, our PINN-based framework is robust enough to start from random initial guesses for the polynomial coefficients of $D_i^*(N)$. Despite the high dimensionality of the parameter space, the optimizer successfully converges to physically meaningful solutions without needing hand-tuned or educated starting values. This flexibility significantly simplifies the deployment of the model, making it broadly applicable even in scenarios where no prior diffusivity estimates (e.g., from binary or ternary systems) are available.

*(iii) Handling High-Dimensional Spaces*

In a five-component alloy system, capturing the full set of composition-dependent tracer diffusivities requires estimating a large number of parameters—typically $15$ polynomial coefficients per element, resulting in a total of $75$ design variables. This high-dimensional optimization problem poses a significant challenge, especially when compounded by noisy experimental data or strong inter-element interactions. To manage this complexity, we employ a moderate-sized neural network architecture with suitable regularization and, where appropriate, staged (layer-wise) training. These strategies help stabilize convergence, mitigate overfitting, and ensure well-conditioned optimization landscapes.

*(iv) Hyperparameter Optimization via Bayesian Strategy*

To further enhance training performance and generalization, we apply Gaussian Process (GP)-based Bayesian optimization to systematically tune both architectural and training-related hyperparameters of the PINN. The search space includes the number of dense layers 1–8, the number of neurons per layer 16–128, activation function (restricted to ($tanh$), and the learning rate $1 \times 10^{-5}$ to $5 \times 10^{-3}$. The optimization was carried out over 100,000 training iterations using the Adam optimizer, which adaptively adjusts step sizes for each parameter based on gradient history. The best-performing configuration was found to be a 6-layer neural network with 22 neurons per layer and a learning rate of $2.5 \times 10^{-5}$. This automated tuning framework enables efficient exploration of the hyperparameter space and contributes significantly to both convergence speed and model robustness.

(v) Experimentally Derived Constraints

Relying solely on composition profiles for numerical optimization can lead to multiple, equally plausible solutions that fail to reflect real-world diffusion behaviors. Direct experimental measurements—such as tracer or impurity diffusivities at the Kirkendall marker plane—provide powerful additional constraints. By incorporating these values into the constraint loss term ($\mathcal{L}_c$), the optimization is "pulled" toward physically meaningful minima, thereby avoiding erroneous mobility coefficients. In other words, a profile-only fit might superficially match the overall concentration gradient yet generate unphysical diffusion parameters, whereas adding even a handful of precise tracer or impurity data points strongly anchors the solution, reducing the risk of converging on a nonphysical set of parameters.

*(vi) Potential Model Enhancements by Adaptive Loss Balancing*

The solution might ignore data constraints if one term (e.g., PDE residual) dominates the early training phase. Methods that dynamically re-scale $\omega_s, \omega_b, \omega_d, \omega_c$ based on gradient magnitudes can yield more robust convergence.

A schematic representation of the constraint-enhanced PINN method flow chart is given in Fig. 1.

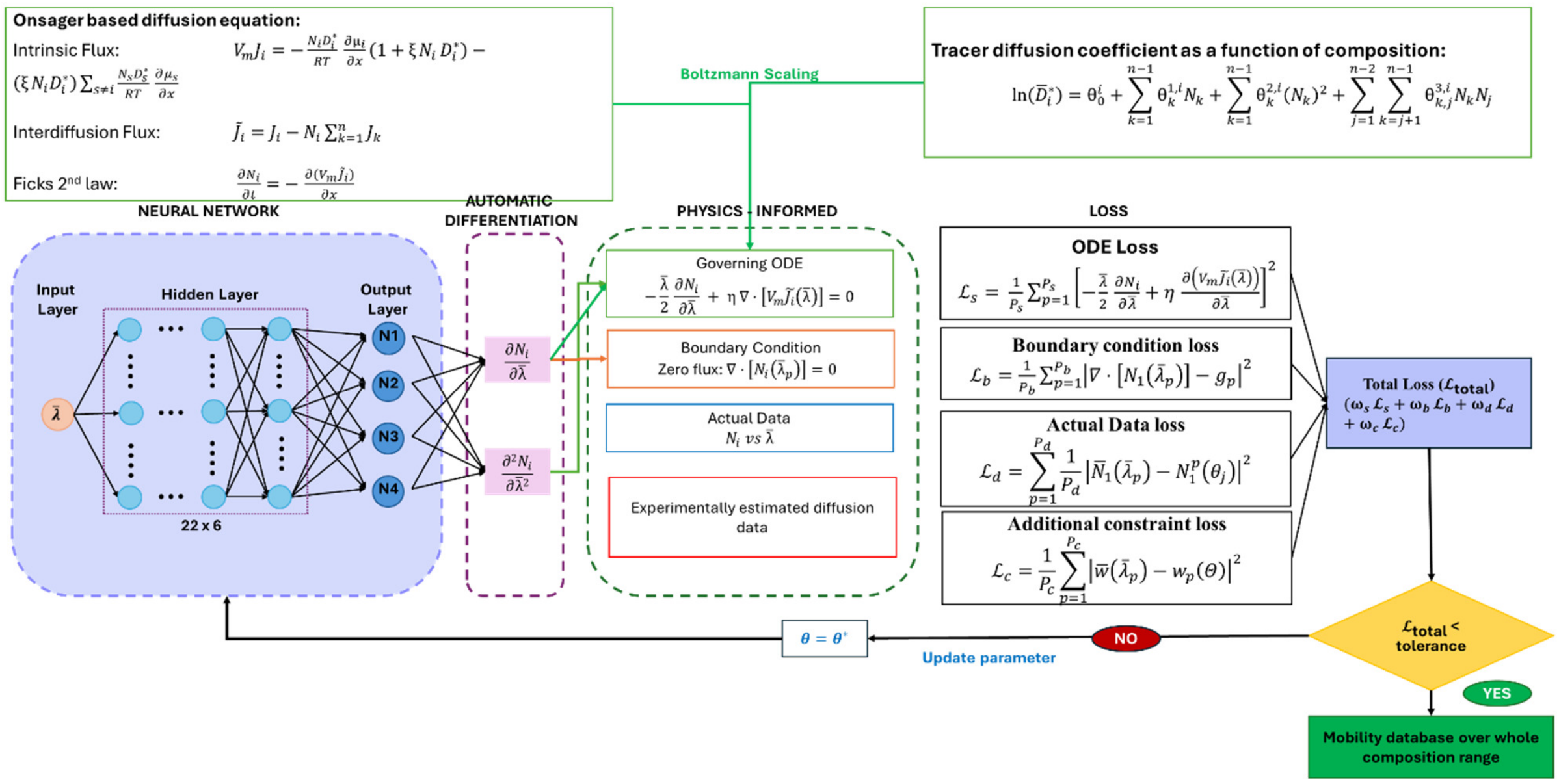

Figure 1 A schematic representation of a constraint-enhanced PINN flow chart.

## 3. Results and discussion

### 3.1 Experimentally estimated diffusion coefficients

Three diffusion couples are produced by coupling Ni, Co and Fe with $(NiCoFeCr)_{85}Mn_{15}$ at 1200 °C by annealing for 50 h. The measured composition profiles of the elements are shown in Fig. 2. The locations of the Kirkendall marker planes are identified by the presence of inert markers along the Kirkendall marker plane, which are shown by the dashed line in the composition profiles. The intrinsic fluxes are calculated at this plane. The diffusion couples are designed strategically in such a way that in DC1, the composition of all the elements except Ni has zero composition ($N_i^- = 0$) in one of the end members *i.e.* by coupling Ni with $(NiCoFeCr)_{85}Mn_{15}$ alloy. Therefore, the equation, as expressed in Eq. 1 for calculating the intrinsic fluxes of these elements (which have zero composition at one end, for example, Co, Fe, Cr and Mn in DC1), reduces to $V_mJ_i = -\frac{1}{2t}\left[N_i^+ \int_{x^-}^{x_K} Y_i dx\right]$. This reduces the error in the calculation of data. On the other hand, the calculation of the intrinsic flux of Ni in DC1, which is not zero at both end members, is prone to induce higher error. A detailed description and analysis of such error analysis is covered in Ref. [11]. Other two diffusion couples, DC2 and DC3, are produced by coupling Co and Fe with the same alloy following a similar strategy.

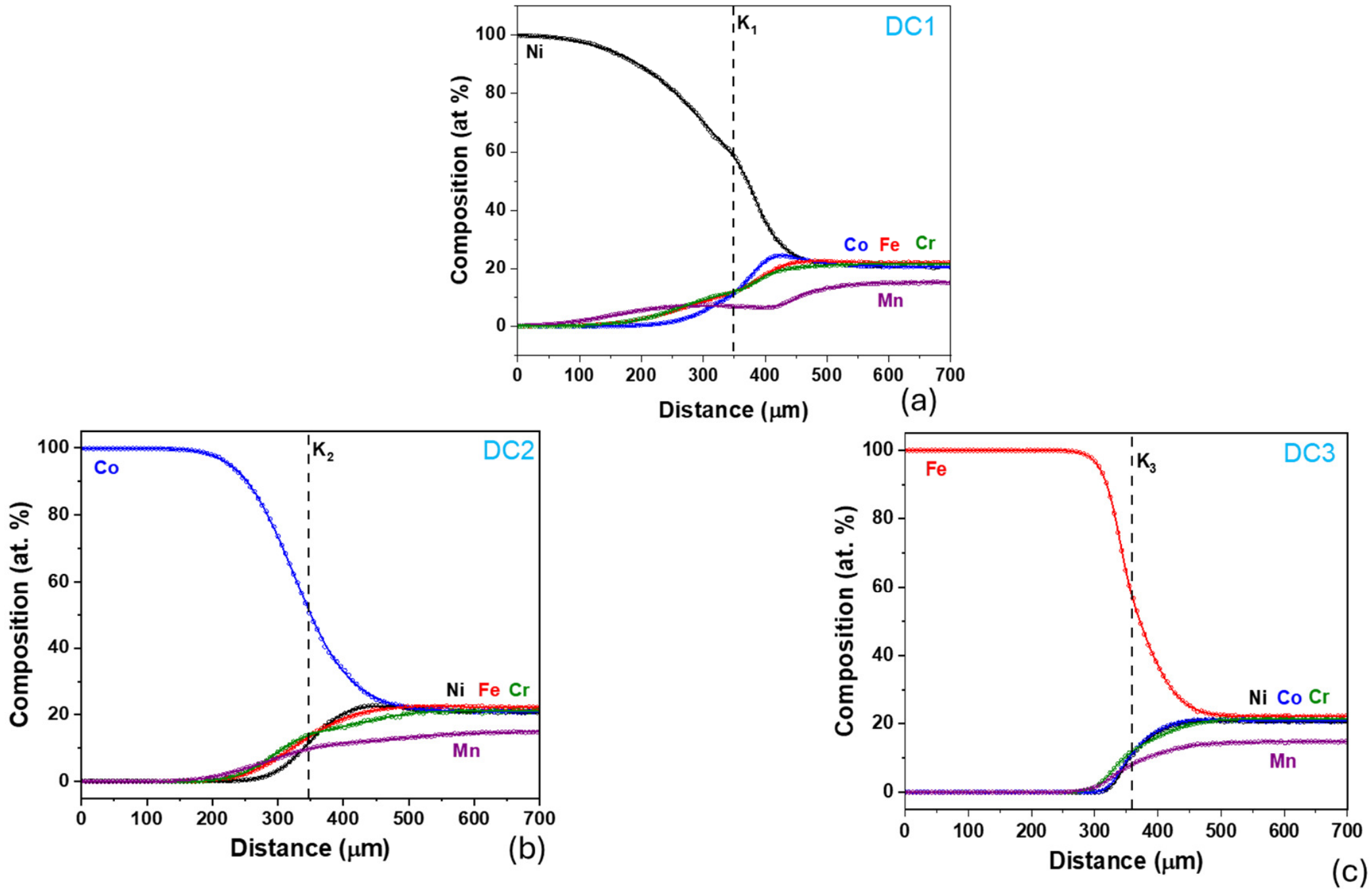

Figure 2 The Diffusion couples produced by coupling (a) DC1: Ni and $(NiCoFeCr)_{85}Mn_{15}$ (b) DC2: Co and $(NiCoFeCr)_{85}Mn_{15}$ and (c) Fe and $(NiCoFeCr)_{85}Mn_{15}$ at 1200 °C for 50h. $K_1$, $K_2$ and $K_3$ indicate the locations of the Kirkendall marker planes.

Further, we need to extract the thermodynamic factors from the available thermodynamic database. The activity profiles of elements extracted from the ThermoCalc database TCHEA7 are shown in Fig. 3. A nice match in the pattern of the activity and diffusion profile is evident. For example, the uphill diffusion profile of Co in DC1 and Ni in DC2 (refer to Fig. 2) resembles the activity profiles of these elements. After calculating the thermodynamic factors from the extracted activity data, the tracer diffusion coefficients of the elements are calculated at the Kirkendall marker plane from Eq. 4. These are listed in Table 2. A trend in diffusivities can be noticed in having $D_{Ni}^{*} \approx D_{Co}^{*} < D_{Fe}^{*} \approx D_{Cr}^{*} < D_{Mn}^{*}$. This falls in the line of estimated or calculated tracer diffusion coefficient in Ni-Co-Fe [11] NiCoFeCr [6,13, 36], NiCoFeCrMn [33, 36] and NiCoFeCrAl [12] systems. Moreover, the diffusivities of all the elements are found to be higher on the Ni-rich side, estimated from DC1 and lower on the Fe-rich side, estimated from DC3. Such a trend of diffusivities can even be found in the impurity and self-diffusion coefficients in pure elements, as listed in Table 3, which are taken from various references [51-59]. Moreover, the impurity diffusion coefficients in pure elements are not much different when the data are calculated by coupling that pure element with the $(NiCoFeCr)_{85}Mn_{15}$ alloy since the location of the marker plane is closer to the pure element. This is consistent with the data calculated in all the diffusion couples, which corroborates the good quality of diffusion coefficients estimated in this study.

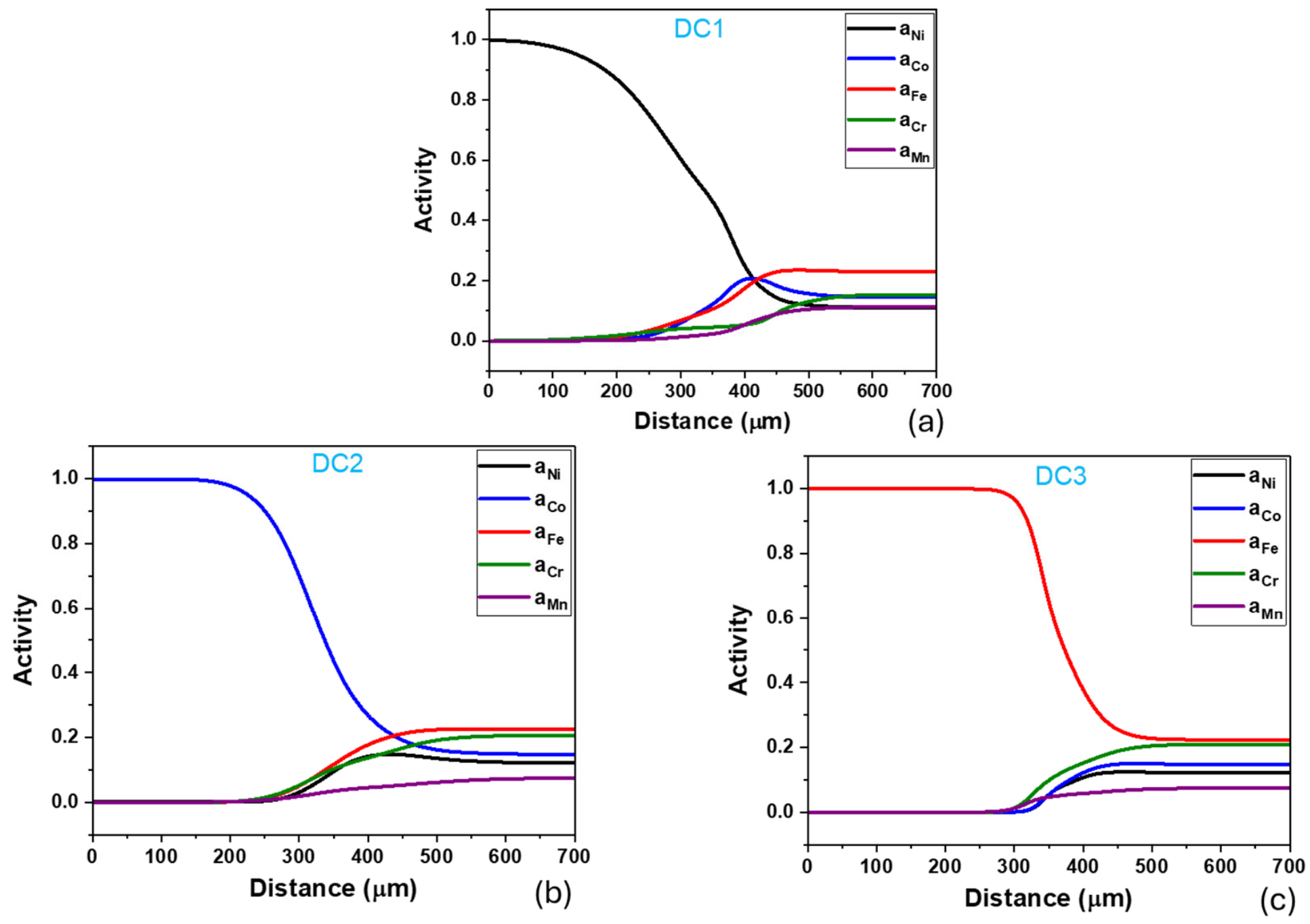

Figure 3 The activity profiles along the diffusion profiles of (a) DC1, (b) DC2, and (c) DC3 extracted from ThermoCalc database TCHEA7 at 1200ºC.

| Marker Plane location | Tracer diffusion coefficient × 10⁻¹⁴ (m²/s) | | | | |
|---|---|---|---|---|---|
| | $D^*_{Ni}$ | $D^*_{Co}$ | $D^*_{Fe}$ | $D^*_{Cr}$ | $D^*_{Mn}$ |
| $K_1$ (DC1)<br>Ni=58.6 Co=11.1 Fe=11.7, Cr= 11.8 Mn=6.8 | 1.25 | 1.16 | 3.0 | 3.2 | 10.8 |
| $K_2$ (DC2)<br>Ni=11.6 Co=51.3 Fe=13.4, Cr= 13.9 Mn=9.8 | 0.59 | 0.69 | 1.2 | 1.9 | 2.6 |
| $K_3$(DC3)<br>Ni=11 Co=11.2 Fe=57.8, Cr= 11.8 Mn=8.1 | 0.28 | 0.26 | 0.63 | 0.83 | 1.9 |

Table. 2 The tracer diffusion coefficients calculated at the Kirkendall marker plane compositions of the diffusion couples at 1200 ºC.

| Pure metal side | Impurity and self diffusion coefficient × 10^-14 (m²/s) | | | | |
|---|---|---|---|---|---|
| | $D^*_{Ni}$ | $D^*_{Co}$ | $D^*_{Fe}$ | $D^*_{Cr}$ | $D^*_{Mn}$ |
| Pure Ni end | 1.46 | 1.62 | 2.57 | 2.64 | 8.2 |
| Pure Co end | 0.42 | 0.44 | 1.1 | 1.3 | 2.1 |
| Pure Fe end | 0.21 | 0.26 | 0.38 | 0.71 | 0.84 |

Table. 3 The impurity diffusion coefficients at 1200 °C extracted from literature [50-59].

From the estimated tracer diffusion coefficients, the intrinsic and interdiffusion coefficients are calculated at the Kirkendall marker composition utilizing Eq. 3 and Eq. 5, respectively. These are listed in Tables 4, 5 and 6, calculated at $K_1$ (of DC1), $K_2$ (of DC2) and $K_3$ (of DC3). The data are calculated neglecting and considering the vacancy wind effect (i.e. by neglecting and considering $\left(1+W_{ij}^{n}\right)$ of Eq. 3) to understand the influence of the cross-phenomenological constants of Onsager formalism. Let us consider the diffusion coefficients calculated from DC1 as listed in Table 4. We can pick the cross-intrinsic diffusion coefficients, which have significant values compared to the main intrinsic diffusion coefficients. The vacancy wind effects of the main intrinsic diffusion coefficients are within the range of 1.01 to 1.09, in which a value of one indicates zero influence of the vacancy wind effect and a higher deviation from one indicates a stronger influence of the vacancy wind effect. Certain cross-intrinsic diffusion coefficients, such as $D_{NiFe}^{Ni}$, $D_{NiCr}^{Ni}$, $D_{NiMn}^{Ni}$, $D_{FeMn}^{Ni}$, $D_{CrMn}^{Ni}$, $D_{MnFe}^{Ni}$ etc., have a significant role in the vacancy wind effects, which are not negligible compared to the main intrinsic diffusion coefficients. Therefore, the influence of this effect when calculated at a relatively high composition of alloying elements should not be neglected. Similar discussions were also reported recently [11, 12]. The vacancy wind effects on interdiffusion coefficients $\left(1+\widetilde{W}_{ij}^{n}\right)$ are found to be, in general, smaller than the influence on the intrinsic diffusion coefficients $\left(1+W_{ij}^{n}\right)$ since the interdiffusion coefficients are a kind of average of *n* intrinsic diffusion coefficients in a n-component system, which could be understood from Eq. 5. Sometimes the vacancy wind effects on intrinsic diffusion indicated from $\left(1+W_{ij}^{n}\right)$ values and interdiffusion coefficients indicated from $\left(1+\widetilde{W}_{ij}^{n}\right)$ show opposite effects with greater or less than one (for example $D_{CoFe}^{Ni}$ and $\widetilde{D}_{CoFe}^{Ni}$ or $D_{CoCr}^{Ni}$ and $\widetilde{D}_{CoCr}^{Ni}$ etc. in Table 4) Therefore, this should not be judged from the interdiffusion coefficients.

Another important point to be noted here. Mostly, interdiffusion coefficients are estimated or attempted to calculate in ternary and multicomponent systems. Following the diffusion mechanisms, the diffusional interactions are discussed based on the signs and values of cross interdiffusion coefficients relative to the values of the main interdiffusion coefficients. Here, we need to question whether the estimated interdiffusion coefficients correctly highlight the diffusional interactions of elements. Let us consider the cross intrinsic diffusion coefficients of Co, such as $D_{CoFe}^{Ni} = -0.04 \times 10^{-14}$, $D_{CoCr}^{Ni} = -0.07 \times 10^{-14}$ and $D_{CoMn}^{Ni} = 0.15 \times 10^{-14}\ m^2/s$ considering the vacancy wind effect (refer Table 4). These are relatively small compared to the main intrinsic diffusion coefficients of Co, *i.e.* $D_{CoCo}^{Ni} = 1.12 \times 10^{-14}\ m^2/s$. Therefore, we can state that Co has very small negative diffusional interactions with Fe and Cr and small positive diffusional interactions with Mn. On the other hand, if we discuss the same considering the interdiffusion coefficients, we have $\tilde{D}_{CoFe}^{Ni} = -0.65 \times 10^{-14}$, $\tilde{D}_{CoCr}^{Ni} = -0.85 \times 10^{-14}$ and $\tilde{D}_{CoMn}^{Ni} = -1.85 \times 10^{-14} m^2/s$. Therefore, it would indicate that Co has high negative diffusional interactions with Fe and Cr, which is higher than 0.5 times of main interdiffusion coefficient of Co. Moreover, Co has very high negative diffusional interaction with Mn ( $\tilde{D}_{CoMn}^{Ni} = -1.85 \times 10^{-15}\ m^2/s$), which has a value even higher than the main interdiffusion coefficient ($\tilde{D}_{CoCo}^{Ni} = 1.04 \times 10^{-15}\ m^2/s$). Therefore, the discussion based on the interdiffusion coefficient is very different compared to the discussion with intrinsic diffusion coefficients. It should be realized that the interdiffusion coefficient is a kind of average of *n* intrinsic diffusion coefficients. For example, following Eq. 5, $\tilde{D}_{CoMn}^{Ni} = (1 - N_{Co})D_{CoMn}^{Ni} - N_{Co}\left(D_{NiMn}^{Ni} + D_{FeMn}^{Ni} + D_{CrMn}^{Ni} + D_{MnMn}^{Ni}\right)$. This leads to a value of $\tilde{D}_{CoMn}^{Ni} = -1.85 \times 10^{-15}\ m^2/s$, which is much higher than the value of $D_{CoMn}^{Ni}$ with an opposite sign because of contributions from other intrinsic diffusion coefficients with higher values. So, the small positive diffusional interaction of Co with Mn is clear from the intrinsic diffusion coefficients is indicated wrongly as a very high negative diffusional interaction if discussed considering the interdiffusion coefficients. Therefore, commenting on the diffusional interactions of an element based on the interdiffusion coefficients is wrong since this does not reflect on the diffusional interactions of a particular element, which is reflected correctly by the knowledge of the intrinsic diffusion coefficients. A similar discussion is valid for the data estimated at other Kirkendall marker planes (K2 and K3) estimated at different compositions, as listed in Tables 5 and 6.

| $\boldsymbol{\phi}_{ij}^{n}$ | | $D_{ij}$ ($\times 10^{-14}$ m$^2$/s) | | | $1+W_{ij}$ | $\tilde{D}_{ij}$ ($\times 10^{-14}$ m$^2$/s) | | | $1+\tilde{W}_{ij}$ |
|---|---|---|---|---|---|---|---|---|---|
| | | | Without VWE | With VWE | | | Without VWE | With VWE | |
| $\boldsymbol{\phi}_{NiCo}^{Ni}$ | -0.1903 | $D_{NiCo}^{Ni}$ | -1.26 | -1.21 | 0.96 | | | | |
| $\boldsymbol{\phi}_{NiFe}^{Ni}$ | -0.2893 | $D_{NiFe}^{Ni}$ | -1.81 | -1.43 | 0.79 | | | | |
| $\boldsymbol{\phi}_{NiCr}^{Ni}$ | -0.3291 | $D_{NiCr}^{Ni}$ | -2.04 | -1.56 | 0.76 | | | | |
| $\boldsymbol{\phi}_{NiMn}^{Ni}$ | -0.2265 | $D_{NiMn}^{Ni}$ | -2.44 | -1.19 | 0.49 | | | | |
| $\boldsymbol{\phi}_{CoCo}^{Ni}$ | 0.9565 | $D_{CoCo}^{Ni}$ | 1.11 | 1.12 | 1.01 | $\tilde{D}_{CoCo}^{Ni}$ | 1.05 | 1.04 | 0.99 |
| $\boldsymbol{\phi}_{CoFe}^{Ni}$ | -0.0977 | $D_{CoFe}^{Ni}$ | -0.11 | -0.04 | 0.38 | $\tilde{D}_{CoFe}^{Ni}$ | -0.58 | -0.65 | 1.11 |
| $\boldsymbol{\phi}_{CoCr}^{Ni}$ | -0.139 | $D_{CoCr}^{Ni}$ | -0.15 | -0.07 | 0.44 | $\tilde{D}_{CoCr}^{Ni}$ | -0.76 | -0.85 | 1.11 |
| $\boldsymbol{\phi}_{CoMn}^{Ni}$ | -0.0379 | $D_{CoMn}^{Ni}$ | -0.07 | 0.15 | - | $\tilde{D}_{CoMn}^{Ni}$ | -1.63 | -1.85 | 1.13 |
| $\boldsymbol{\phi}_{FeCo}^{Ni}$ | -0.0091 | $D_{FeCo}^{Ni}$ | -0.03 | -0.01 | 0.18 | $\tilde{D}_{FeCo}^{Ni}$ | -0.09 | -0.09 | 1.00 |
| $\boldsymbol{\phi}_{FeFe}^{Ni}$ | 1.1513 | $D_{FeFe}^{Ni}$ | 3.45 | 3.64 | 1.05 | $\tilde{D}_{FeFe}^{Ni}$ | 2.95 | 2.99 | 1.01 |
| $\boldsymbol{\phi}_{FeCr}^{Ni}$ | 0.1481 | $D_{FeCr}^{Ni}$ | 0.44 | 0.68 | 1.53 | $\tilde{D}_{FeCr}^{Ni}$ | -0.21 | -0.15 | 0.74 |
| $\boldsymbol{\phi}_{FeMn}^{Ni}$ | 0.1427 | $D_{FeMn}^{Ni}$ | 0.74 | 1.33 | 1.81 | $\tilde{D}_{FeMn}^{Ni}$ | -0.92 | -0.78 | 0.85 |
| $\boldsymbol{\phi}_{CrCo}^{Ni}$ | -0.0118 | $D_{CrCo}^{Ni}$ | -0.04 | -0.01 | 0.37 | $\tilde{D}_{CrCo}^{Ni}$ | -0.11 | -0.10 | 0.94 |
| $\boldsymbol{\phi}_{CrFe}^{Ni}$ | 0.1846 | $D_{CrFe}^{Ni}$ | 0.59 | 0.79 | 1.33 | $\tilde{D}_{CrFe}^{Ni}$ | 0.08 | 0.14 | 1.62 |
| $\boldsymbol{\phi}_{CrCr}^{Ni}$ | 1.3389 | $D_{CrCr}^{Ni}$ | 4.27 | 4.52 | 1.06 | $\tilde{D}_{CrCr}^{Ni}$ | 3.62 | 3.69 | 1.02 |
| $\boldsymbol{\phi}_{CrMn}^{Ni}$ | 0.2413 | $D_{CrMn}^{Ni}$ | 1.33 | 1.97 | 1.48 | $\tilde{D}_{CrMn}^{Ni}$ | -0.33 | -0.16 | 0.48 |
| $\boldsymbol{\phi}_{MnCo}^{Ni}$ | 0.1168 | $D_{MnCo}^{Ni}$ | 0.78 | 0.82 | 1.06 | $\tilde{D}_{MnCo}^{Ni}$ | 0.74 | 0.78 | 1.05 |
| $\boldsymbol{\phi}_{MnFe}^{Ni}$ | 0.3464 | $D_{MnFe}^{Ni}$ | 2.17 | 2.56 | 1.18 | $\tilde{D}_{MnFe}^{Ni}$ | 1.88 | 2.18 | 1.16 |
| $\boldsymbol{\phi}_{MnCr}^{Ni}$ | 0.4821 | $D_{MnCr}^{Ni}$ | 3.00 | 3.49 | 1.16 | $\tilde{D}_{MnCr}^{Ni}$ | 2.63 | 3.01 | 1.15 |
| $\boldsymbol{\phi}_{MnMn}^{Ni}$ | 1.3467 | $D_{MnMn}^{Ni}$ | 14.54 | 15.79 | 1.09 | $\tilde{D}_{MnMn}^{Ni}$ | 13.58 | 14.56 | 1.07 |

Table 4 Tracer, intrinsic and interdiffusion coefficients calculated at $K_1$ (DC1) at 1200°C. VWE stand for vacancy wind effect.

| $\boldsymbol{\phi}_{ij}^{n}$ | | $D_{ij}$ ($\times 10^{-14}$ m$^2$/s) | | | $1+W_{ij}$ | $\widetilde{D}_{ij}$ ($\times 10^{-14}$ m$^2$/s) | | | $1+\widetilde{W}_{ij}$ |
|---|---|---|---|---|---|---|---|---|---|
| | | | Without VWE | With VWE | | | Without VWE | With VWE | |
| $\boldsymbol{\phi}_{NiCo}^{Ni}$ | -4.2927 | $D_{NiCo}^{Ni}$ | -0.57 | -0.57 | 1.00 | | | | |
| $\boldsymbol{\phi}_{NiFe}^{Ni}$ | -1.1939 | $D_{NiFe}^{Ni}$ | -0.61 | -0.59 | 0.97 | | | | |
| $\boldsymbol{\phi}_{NiCr}^{Ni}$ | -1.3078 | $D_{NiCr}^{Ni}$ | -0.64 | -0.6 | 0.94 | | | | |
| $\boldsymbol{\phi}_{NiMn}^{Ni}$ | -1.0318 | $D_{NiMn}^{Ni}$ | -0.72 | -0.67 | 0.93 | | | | |
| $\boldsymbol{\phi}_{CoCo}^{Ni}$ | 0.924 | $D_{CoCo}^{Ni}$ | 0.64 | 0.66 | 1.03 | $\widetilde{D}_{CoCo}^{Ni}$ | 0.51 | 0.5 | 0.98 |
| $\boldsymbol{\phi}_{CoFe}^{Ni}$ | -0.1171 | $D_{CoFe}^{Ni}$ | -0.31 | -0.19 | 0.61 | $\widetilde{D}_{CoFe}^{Ni}$ | -0.99 | -1.06 | 1.07 |
| $\boldsymbol{\phi}_{CoCr}^{Ni}$ | -0.1454 | $D_{CoCr}^{Ni}$ | -0.37 | -0.17 | 0.46 | $\widetilde{D}_{CoCr}^{Ni}$ | -1.54 | -1.66 | 1.08 |
| $\boldsymbol{\phi}_{CoMn}^{Ni}$ | -0.1261 | $D_{CoMn}^{Ni}$ | -0.46 | -0.16 | 0.35 | $\widetilde{D}_{CoMn}^{Ni}$ | -2.15 | -2.32 | 1.08 |
| $\boldsymbol{\phi}_{FeCo}^{Ni}$ | -0.1598 | $D_{FeCo}^{Ni}$ | -0.05 | -0.04 | 0.80 | $\widetilde{D}_{FeCo}^{Ni}$ | -0.08 | -0.08 | 1.00 |
| $\boldsymbol{\phi}_{FeFe}^{Ni}$ | 1.0785 | $D_{FeFe}^{Ni}$ | 1.29 | 1.35 | 1.05 | $\widetilde{D}_{FeFe}^{Ni}$ | 1.12 | 1.12 | 1.00 |
| $\boldsymbol{\phi}_{FeCr}^{Ni}$ | 0.0756 | $D_{FeCr}^{Ni}$ | 0.09 | 0.18 | 2.00 | $\widetilde{D}_{FeCr}^{Ni}$ | -0.22 | -0.21 | 0.95 |
| $\boldsymbol{\phi}_{FeMn}^{Ni}$ | 0.1052 | $D_{FeMn}^{Ni}$ | 0.17 | 0.31 | 1.82 | $\widetilde{D}_{FeMn}^{Ni}$ | -0.27 | -0.26 | 0.96 |
| $\boldsymbol{\phi}_{CrCo}^{Ni}$ | -0.0051 | $D_{CrCo}^{Ni}$ | Neg. | 0.01 | - | $\widetilde{D}_{CrCo}^{Ni}$ | -0.04 | -0.03 | 0.75 |
| $\boldsymbol{\phi}_{CrFe}^{Ni}$ | 0.1356 | $D_{CrFe}^{Ni}$ | 0.27 | 0.36 | 1.33 | $\widetilde{D}_{CrFe}^{Ni}$ | 0.08 | 0.12 | 1.50 |
| $\boldsymbol{\phi}_{CrCr}^{Ni}$ | 1.2016 | $D_{CrCr}^{Ni}$ | 2.28 | 2.44 | 1.07 | $\widetilde{D}_{CrCr}^{Ni}$ | 1.97 | 2.03 | 1.03 |
| $\boldsymbol{\phi}_{CrMn}^{Ni}$ | 0.2378 | $D_{CrMn}^{Ni}$ | 0.64 | 0.86 | 1.34 | $\widetilde{D}_{CrMn}^{Ni}$ | 0.19 | 0.28 | 1.47 |
| $\boldsymbol{\phi}_{MnCo}^{Ni}$ | 0.4666 | $D_{MnCo}^{Ni}$ | 0.23 | 0.25 | 1.09 | $\widetilde{D}_{MnCo}^{Ni}$ | 0.21 | 0.22 | 1.05 |
| $\boldsymbol{\phi}_{MnFe}^{Ni}$ | 0.3628 | $D_{MnFe}^{Ni}$ | 0.69 | 0.77 | 1.12 | $\widetilde{D}_{MnFe}^{Ni}$ | 0.56 | 0.61 | 1.09 |
| $\boldsymbol{\phi}_{MnCr}^{Ni}$ | 0.501 | $D_{MnCr}^{Ni}$ | 0.91 | 1.06 | 1.16 | $\widetilde{D}_{MnCr}^{Ni}$ | 0.69 | 0.78 | 1.13 |
| $\boldsymbol{\phi}_{MnMn}^{Ni}$ | 1.4045 | $D_{MnMn}^{Ni}$ | 3.65 | 3.86 | 1.06 | $\widetilde{D}_{MnMn}^{Ni}$ | 3.33 | 3.45 | 1.04 |

Table 5 Tracer, intrinsic and interdiffusion coefficients calculated at $K_2$ (DC2) at 1200°C. VWE stands for vacancy wind effect.

| $\boldsymbol{\phi}_{ij}^{n}$ | | $D_{ij}$ ($\times$ 10$^{-14}$ m$^2$/s) | | | 1 + $W_{ij}$ | $\widetilde{D}_{ij}$ ($\times$ 10$^{-14}$ m$^2$/s) | | | 1 + $\widetilde{W}_{ij}$ |
|---|---|---|---|---|---|---|---|---|---|
| | | | Without VWE | With VWE | | | Without VWE | With VWE | |
| $\boldsymbol{\phi}_{NiCo}^{Ni}$ | -0.9435 | $D_{NiCo}^{Ni}$ | -0.26 | -0.26 | 1.00 | | | | |
| $\boldsymbol{\phi}_{NiFe}^{Ni}$ | -5.3327 | $D_{NiFe}^{Ni}$ | -0.28 | -0.27 | 0.96 | | | | |
| $\boldsymbol{\phi}_{NiCr}^{Ni}$ | -1.1488 | $D_{NiCr}^{Ni}$ | -0.29 | -0.28 | 0.97 | | | | |
| $\boldsymbol{\phi}_{NiMn}^{Ni}$ | -0.9285 | $D_{NiMn}^{Ni}$ | -0.35 | -0.32 | 0.91 | | | | |
| $\boldsymbol{\phi}_{CoCo}^{Ni}$ | 0.9924 | $D_{CoCo}^{Ni}$ | 0.25 | 0.25 | 1.00 | $\widetilde{D}_{CoCo}^{Ni}$ | 0.25 | 0.24 | 0.96 |
| $\boldsymbol{\phi}_{CoFe}^{Ni}$ | -0.6128 | $D_{CoFe}^{Ni}$ | -0.03 | -0.02 | 0.67 | $\widetilde{D}_{CoFe}^{Ni}$ | -0.1 | -0.12 | 1.20 |
| $\boldsymbol{\phi}_{CoCr}^{Ni}$ | -0.1681 | $D_{CoCr}^{Ni}$ | -0.04 | -0.03 | 0.75 | $\widetilde{D}_{CoCr}^{Ni}$ | -0.15 | -0.17 | 1.13 |
| $\boldsymbol{\phi}_{CoMn}^{Ni}$ | -0.1479 | $D_{CoMn}^{Ni}$ | -0.05 | -0.03 | 0.60 | $\widetilde{D}_{CoMn}^{Ni}$ | -0.29 | -0.34 | 1.17 |
| $\boldsymbol{\phi}_{FeCo}^{Ni}$ | -0.0286 | $D_{FeCo}^{Ni}$ | -0.09 | -0.08 | 0.89 | $\widetilde{D}_{FeCo}^{Ni}$ | -0.14 | -0.14 | 1.00 |
| $\boldsymbol{\phi}_{FeFe}^{Ni}$ | 0.9122 | $D_{FeFe}^{Ni}$ | 0.57 | 0.67 | 1.18 | $\widetilde{D}_{FeFe}^{Ni}$ | 0.2 | 0.19 | 0.95 |
| $\boldsymbol{\phi}_{FeCr}^{Ni}$ | -0.0245 | $D_{FeCr}^{Ni}$ | -0.08 | 0.07 | -0.88 | $\widetilde{D}_{FeCr}^{Ni}$ | -0.63 | -0.64 | 1.02 |
| $\boldsymbol{\phi}_{FeMn}^{Ni}$ | 0.0087 | $D_{FeMn}^{Ni}$ | 0.04 | 0.37 | 9.25 | $\widetilde{D}_{FeMn}^{Ni}$ | -1.21 | -1.24 | 1.02 |
| $\boldsymbol{\phi}_{CrCo}^{Ni}$ | -0.0187 | $D_{CrCo}^{Ni}$ | -0.02 | -0.01 | 0.50 | $\widetilde{D}_{CrCo}^{Ni}$ | -0.03 | -0.03 | 1.00 |
| $\boldsymbol{\phi}_{CrFe}^{Ni}$ | 0.1373 | $D_{CrFe}^{Ni}$ | 0.02 | 0.05 | 2.50 | $\widetilde{D}_{CrFe}^{Ni}$ | -0.05 | -0.05 | 1.00 |
| $\boldsymbol{\phi}_{CrCr}^{Ni}$ | 1.1017 | $D_{CrCr}^{Ni}$ | 0.91 | 0.95 | 1.04 | $\widetilde{D}_{CrCr}^{Ni}$ | 0.8 | 0.81 | 1.01 |
| $\boldsymbol{\phi}_{CrMn}^{Ni}$ | 0.0911 | $D_{CrMn}^{Ni}$ | 0.11 | 0.2 | 1.82 | $\widetilde{D}_{CrMn}^{Ni}$ | -0.15 | -0.13 | 0.87 |
| $\boldsymbol{\phi}_{MnCo}^{Ni}$ | 0.1386 | $D_{MnCo}^{Ni}$ | 0.19 | 0.2 | 1.05 | $\widetilde{D}_{MnCo}^{Ni}$ | 0.18 | 0.19 | 1.06 |
| $\boldsymbol{\phi}_{MnFe}^{Ni}$ | 1.3701 | $D_{MnFe}^{Ni}$ | 0.36 | 0.41 | 1.14 | $\widetilde{D}_{MnFe}^{Ni}$ | 0.31 | 0.34 | 1.10 |
| $\boldsymbol{\phi}_{MnCr}^{Ni}$ | 0.3501 | $D_{MnCr}^{Ni}$ | 0.45 | 0.51 | 1.13 | $\widetilde{D}_{MnCr}^{Ni}$ | 0.38 | 0.41 | 1.08 |
| $\boldsymbol{\phi}_{MnMn}^{Ni}$ | 1.2709 | $D_{MnMn}^{Ni}$ | 2.41 | 2.55 | 1.06 | $\widetilde{D}_{MnMn}^{Ni}$ | 2.24 | 2.33 | 1.04 |

Table 6 Tracer, intrinsic and interdiffusion coefficients calculated at $K_3$ (DC3) at 1200°C. VWE stands for vacancy wind effect.

Countering a similar discussion and comment made by us in another system on the importance of intrinsic diffusion coefficients for discussing the diffusional interactions instead of based on interdiffusion coefficients [11], Nayak and Kulkarni [60] commented this is incorrect stating both intrinsic and interdiffusion coefficients are important parameters for a complete understanding multicomponent diffusion process since both are phenomenological constants measured in different frames of reference, functions of composition through the tracer diffusion coefficients and thermodynamic factors. First of all, we are aware of these basic fundamental details. Their comment rather wrongly indicates as if we are stating against the importance of all types of diffusion coefficients. This is not correct. We are saying categorically that intrinsic diffusion coefficients only should be used for the discussion on diffusional interactions in concentrated or medium and high alloys. The interdiffusion coefficients are definitely important parameters to correlate/calculate the diffusion profiles but can be misleading if these are used for discussion of diffusional interactions in such alloys, which is done. for example, in Ref. [5, 61]. Both their [5] and our study [6] at a composition close to equiatomic in the Ni-Co-Fe-Cr system (although conducted at different temperatures) show that the cross to main interdiffusion coefficients of Ni and Co, i.e. $\frac{\tilde{D}_{NiCo}^{Cr}}{\tilde{D}_{NiNi}^{Cr}}$ and $\frac{\tilde{D}_{CoNi}^{Cr}}{\tilde{D}_{CoCo}^{Cr}}$ to be greater than 0.5. However, our analysis of the intrinsic diffusion coefficient indicates that the cross to main intrinsic diffusion coefficients i.e. $\frac{D_{NiCo}^{Cr}}{D_{NiNi}^{Cr}}$ and $\frac{D_{CoNi}^{Cr}}{D_{CoCo}^{Cr}}$ to be less than 0.25 [6]. Moreover, the values of these interdiffusion coefficients are found to be higher than the intrinsic diffusion coefficients since the intrinsic diffusion coefficients $D_{CrNi}^{Cr}$, $D_{CrCo}^{Cr}$ and $D_{CrFe}^{Cr}$ are significantly higher with opposite sign and included in these interdiffusion coefficients. Therefore, the actual diffusional interactions identified by the intrinsic diffusion coefficients are less than the diffusional interactions indicated by the interdiffusion coefficients. It is true that both the diffusion coefficients are related to tracer diffusion coefficients and thermodynamic factors. However, one should realize that the intrinsic diffusion coefficient signifies the diffusion of a particular element (for example refer to Eq. 3). If we neglect the vacancy wind effect, it is related to one tracer diffusion coefficient and one thermodynamic factor. On the other hand, by replacing Eq. 3 in Eq. 5 (suppose neglecting the vacancy wind effect again), the interdiffusion coefficients are kind of average of *n* tracer and *n* thermodynamic factors in which thermodynamic factors can have different signs (positive and negative). When we consider the vacancy wind effect, this contribution to the intrinsic diffusion coefficient is related to the tracer and thermodynamic factors of all the elements (in the vacancy wind effect part), leading to a complex correlation. Still, it signifies the intrinsic diffusion coefficient of a particular element under the thermodynamic driving force compared to a kind of average value of *n* intrinsic diffusion coefficients represented by the interdiffusion coefficient. Kulkarni et al. [5, 60] could have realized the difference in intrinsic and interdiffusion coefficients by analyzing the data while they calculated back the interdiffusion coefficients [5] from the available tracer diffusion coefficients [33] utilizing the thermodynamic data from the ThermoCalc database, which, they, unfortunately, must have ignored. We have followed the same method of calculation of intrinsic and interdiffusion coefficients from the equation scheme established by Manning utilizing the thermodynamic database of ThermoCalc only to highlight the significance of intrinsic and interdiffusion coefficients [6, 11-13]. The problem with interdiffusion coefficients for discussing the diffusion mechanism can also be understood from their comment on the diffusion studies in the Ni-Cu-Fe system [60]. The reason for the mismatch in the sign of thermodynamic factor $\emptyset_{CuNi}^{Fe}$ and the interdiffusion coefficient $\widetilde{D}_{CuNi}^{Fe}$ is debated by them because of the role of the vacancy wind effect. However, it can be misleading without looking at the correlation between interdiffusion and intrinsic diffusion coefficients, for example, $\widetilde{D}_{CuNi}^{Fe} = (1 - N_{Cu})D_{CuNi}^{Fe} - N_{Cu}(D_{NiNi}^{Fe} + D_{FeNi}^{Fe})$. By substituting Eq. 3, this correlation of interdiffusion coefficient can be expressed with tracer diffusion coefficients of $D_{Ni}^{*}$, $D_{Co}^{*}$, $D_{Fe}^{*}$ and the thermodynamic factors $\emptyset_{CuNi}^{Fe}$, $\emptyset_{NiNi}^{Fe}$ and $\emptyset_{FeNi}^{Fe}$. The tracer diffusion coefficients have positive signs, but thermodynamic factors can have both positive and negative signs. $\emptyset_{FeNi}^{Fe}$ has a negative sign with a higher absolute value compared to other thermodynamic factors, as they have reported [60]. Therefore, interdiffusion coefficient $\widetilde{D}_{CuNi}^{Fe}$ can have the same or opposite sign compared to $\emptyset_{CuNi}^{Fe}$, depending on the values of tracer diffusion coefficients, values and sign of thermodynamic factors at the composition of interest. This would have a similar sign for sure only when calculated at composition $N_{Cu}$ much smaller compared to $(1 - N_{Cu})$ such that the second term in the above equation is negligible. They have mentioned these equation schemes but unfortunately ignored these points of view [60]. The vacancy wind effect can change the sign of intrinsic diffusion coefficient as it is found for $D_{CoMn}^{Ni}$ only in Table 4 in this study, however, the much higher value of $\widetilde{D}_{CoMn}^{Ni}$ compared to $D_{CoMn}^{Ni}$ with a different sign (refer the values considering the vacancy wind effect) clearly indicating the influence of other intrinsic diffusion coefficients. The difference in sign of $\phi_{FeCr}^{Ni}$ and $\widetilde{D}_{FeCr}^{Ni}$, $\phi_{FeMn}^{Ni}$ and $\widetilde{D}_{FeMn}^{Ni}$, $\phi_{CrMn}^{Ni}$ and $\widetilde{D}_{CrMn}^{Ni}$ is also because of the influence of other intrinsic diffusion coefficients since the intrinsic diffusion coefficients and the thermodynamic factors have the same sign even when we consider/neglect the vacancy wind effect. This again clearly indicates the importance of intrinsic diffusion coefficients for the discussion of the diffusional interactions, although interdiffusion coefficients are important for direct correlation or calculation of the diffusion profile. However, inconsistent interdiffusion coefficients estimated by them in Ref. [60] (compared to the influence of unknown error in thermodynamic data, which they have identified as the source of main error) are not suitable for the calculation of reliable intrinsic diffusion coefficients facilitating such discussion, which is discussed in detail in Ref. [10, 23].

It should not be forgotten that even in a binary 1-2 system, the interdiffusion coefficient in a concentrated alloy cannot be said to be the diffusion coefficient of element 1 or 2 as expressed by the intrinsic diffusion coefficients. This is the reason for extensive research available on estimating tracer and intrinsic diffusion coefficients in binary systems. On the other hand, mostly (except a very few) interdiffusion coefficients were only calculated/estimated in ternary systems for a very long time and, recently, even in multicomponent alloys. This was still fine in the alloy with a small composition to discuss the diffusional interactions with interdiffusion coefficients since these values are similar to intrinsic diffusion coefficients. For example, suppose in a $n$-rich alloy of $i - j - n$ system, we have $\tilde{D}_{ij}^{n} \approx D_{ij}^{n}$ because of $N_i, N_j \ll N_n$ (refer to Eq. 5). Therefore, we can discuss the diffusion interactions of $i$ and $j$ with $\tilde{D}_{ij}^{n}$ and $\tilde{D}_{ji}^{n}$. Still, intrinsic diffusion coefficients are important to calculate, without which the diffusional interactions of the major element with minor elements ($D_{ni}^{n}$ and $D_{nj}^{n}$) will not be known [27, 30]. Therefore, in summary, the estimation of only the interdiffusion coefficient is not enough or can be misleading for the discussion of diffusional interactions in concentrated medium and high entropy alloys. Moreover, it is also common practice to discuss the relative mobilities of elements by comparing the main interdiffusion coefficients only. We have shown that such discussion considering different elements as the dependent variable can be very different or opposite, making it confusing when only these parameters are considered for such discussion (depending on the systems). Therefore, these are often termed vague [3], although one can back-calculate the same diffusion profiles considering diffusion coefficients with any element as the dependent variable. Of course, sometimes, especially in intermetallic compounds in the absence of suitable thermodynamic data, it is difficult to estimate the intrinsic diffusion coefficients, and we have no other option but to try to have an impression of the diffusional interactions based on interdiffusion coefficients only. However, we prefer calculating the intrinsic diffusion coefficients in solid solutions when thermodynamic data are available in the recognized databanks. It should be also noted that the calculation of tracer and intrinsic diffusion coefficients is more often sensitive to the quality of the diffusion couples produced compared to the unknown reliability of thermodynamic data available, which is reflected in our several articles [6, 11-13] and also in this study by calculating very consistent and tracer coefficients. A detailed discussion on this is included in Ref. [10].

As already explained, we could estimate the diffusion coefficients only at the Kirkendall marker plane based on experimental analysis. The impurity diffusion coefficients at the pure end members are also known. However, it is important to establish a composition-dependent mobility database over a very broad composition range. This is not easy following the diffusion couple experiments since it needs multiple/numerous diffusion couples to produce, which is not an easy task in a multicomponent space. Therefore, we need to combine it with numerical analysis to extract the diffusion coefficients over the whole composition range of the diffusion couples. As explained next, the experimentally estimated data need to be utilized as an equality constraint, which otherwise may not extract reliable diffusion coefficients. This is demonstrated based on a very advanced PINN method established first time for multicomponent diffusion.

### 3.2 PINN optimization for extracting the diffusion coefficients over the whole composition range of the diffusion couples

The PINN-based numerical inverse method described here uses Boltzmann-transformed ODEs, polynomial expansions for tracer diffusivities, and gradient-based adjoint optimization to reconstruct composition-dependent diffusion coefficients in high-dimensional alloy systems. By embedding both PDE/ODE physics and experimentally derived constraints into a single loss function, one can obtain a mesh-free, open-source alternative to closed-source parameter-optimization tools following the scheme described in section 2.2.

A few additional methodologies for the reliable generation of mobility databases need to be specifically highlighted:

(i) Thermodynamic Integration and Vacancy-Wind Correction: Our method incorporates thermodynamic effects directly into the flux formulation through the chemical potentials $\mu_i$, which naturally capture diffusional interactions in concentrated multicomponent alloys. These composition-dependent driving forces are essential. We have already highlighted the importance of considering the vacancy wind effect proposed by Manning, which has a significant influence on several cross-intrinsic diffusion coefficients. Together, the chemical potential gradients and vacancy-wind corrections form a consistent and physically grounded framework that faithfully captures both thermodynamic and kinetic coupling-critical for simulating diffusion in strongly interacting multicomponent alloys.

(ii) Robustness to Initial Guesses: Unlike general inverse modelling approaches that require careful initialization to ensure convergence, our PINN-based framework is robust enough to start from random initial guesses for the polynomial coefficients of $D_i^*(N)$. Despite the high dimensionality of the parameter space, the optimizer successfully converges to physically meaningful solutions without needing hand-tuned or educated starting values.

(iii) Handling High-Dimensional Spaces: In a five-component alloy system, capturing the full set of composition-dependent tracer diffusivities requires estimating a large number of parameters—typically $15$ polynomial coefficients per element, resulting in a total of $75$ design variables. This high-dimensional optimization problem poses a significant challenge, especially when compounded by noisy experimental data or strong inter-element interactions. To manage this complexity, we employ a moderate-sized neural network architecture with suitable regularization and, where appropriate, staged (layer-wise) training. These strategies help stabilize convergence, mitigate overfitting, and ensure well-conditioned optimization landscapes.

(iv) Hyperparameter Optimization via Bayesian Strategy: To further enhance training performance and generalization, we apply Gaussian Process (GP)-based Bayesian optimization to systematically tune both architectural and training-related hyperparameters of the PINN. The search space includes the number of dense layers 1–8, the number of neurons per layer 16–128, activation function (restricted to $(tanh)$), and the learning rate $1 \times 10^{-5}$ to $5 \times 10^{-3}$. The optimization was carried out over 100,000 training iterations using the Adam optimizer, which adaptively adjusts step sizes for each parameter based on gradient history. The best-performing configuration was found to be a 6-layer neural network with 22 neurons per layer and a learning rate of $2.5 \times 10^{-5}$. This automated tuning framework enables efficient exploration of the hyperparameter space and contributes significantly to both convergence speed and model robustness.

(v) Experimentally Derived Constraints: Relying solely on composition profiles for numerical optimization can lead to multiple, equally plausible solutions that fail to reflect real-world diffusion behaviors. Direct experimental measurements—such as tracer or impurity diffusivities at the Kirkendall marker plane—provide powerful additional constraints. By incorporating these values into the constraint loss term $(\mathcal{L}_c)$, the optimization is “pulled” toward physically meaningful minima, thereby avoiding erroneous mobility coefficients. In other words, a profile-only fit might superficially match the overall concentration gradient yet generate unphysical diffusion parameters, whereas adding even a handful of precise tracer or impurity data points strongly anchors the solution, reducing the risk of converging on a nonphysical set of parameters.

(vi) Potential Model Enhancements by Adaptive Loss Balancing: If one term (e.g., PDE residual) dominates the early training phase, the solution might ignore data constraints. Methods that dynamically re-scale $\omega_s, \omega_b, \omega_d, \omega_c$ (refer to section 2.2.4) based on gradient magnitudes can yield more robust convergence.

It was a common practice to extract the composition-dependent diffusion coefficients by optimizing the diffusion profiles (or interdiffusion flux and composition gradient). However, we noticed that it does not ensure extracting intrinsic coefficients in binary [9] or pseudo-binary diffusion couples [8]. A similar problem is witnessed in this for the extraction of the tracer diffusion coefficient after optimizing the composition profile and thermodynamic parameter. Although a very good match with the composition profile was found, the extracted tracer diffusion coefficients were not reliable in all three diffusion couples produced in this study, as shown in Fig. 4, 5, and 6. Although the predicted composition profiles visually match the experimental data, this is a classic example of an ill-posed inverse problem. Multiple combinations of tracer diffusivities can yield similar concentration profiles, especially in multicomponent systems. This underlines the non-uniqueness of solutions when constraints like experimental tracers or impurity diffusivities are not included. To improve the reliability of the extracted data, first, the tracer diffusion coefficients estimated at the Kirkendall marker plane were used as equality constraints. This improved the optimization significantly by passing the tracer diffusion coefficients through the experimentally estimated data. However, a mismatch was still noticed with the experimentally estimated impurity and self-diffusion coefficients available in the literature, although the match with the diffusion profiles is again excellent. Therefore, the impurity and self-diffusion coefficients were also used as the additional equality constraints in the next step. Since impurity diffusion coefficients cannot be given as the equality constraint at the unaffected end-member composition of pure elements, i.e. at the 100 at. %composition of pure elements Ni, Co and Fe, these were given at a distance related to the composition of almost 0.5 at. % of the alloying elements since the diffusion coefficients are expected to be similar for such a composition.

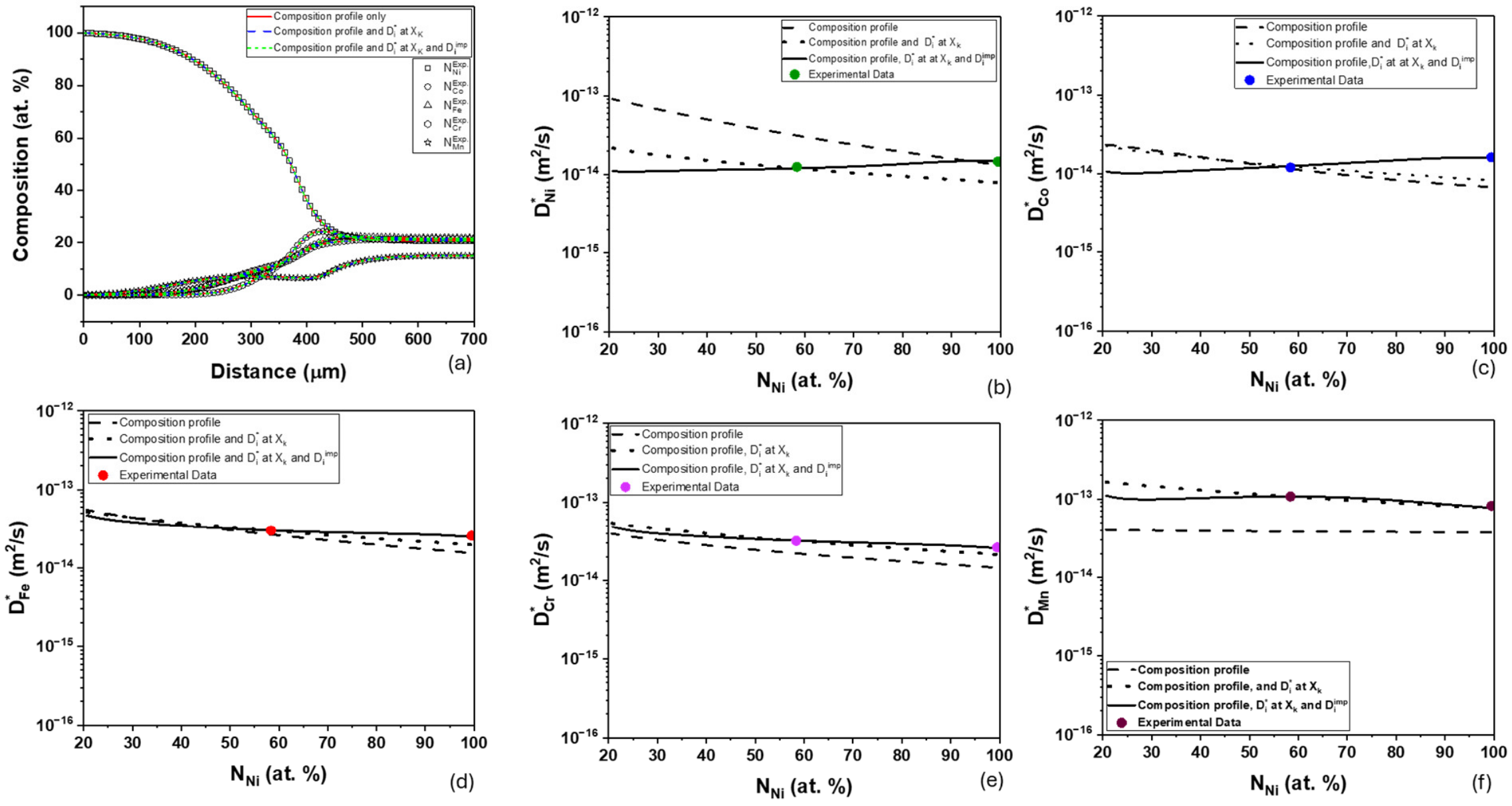

Figure 4 A comparison of PINN optimized (a) diffusion profile of DC1. Composition-dependent tracer diffusion coefficients of (b) Ni, (c) Co, (d) Fe, (e) Cr, (f) Mn with experimental data.

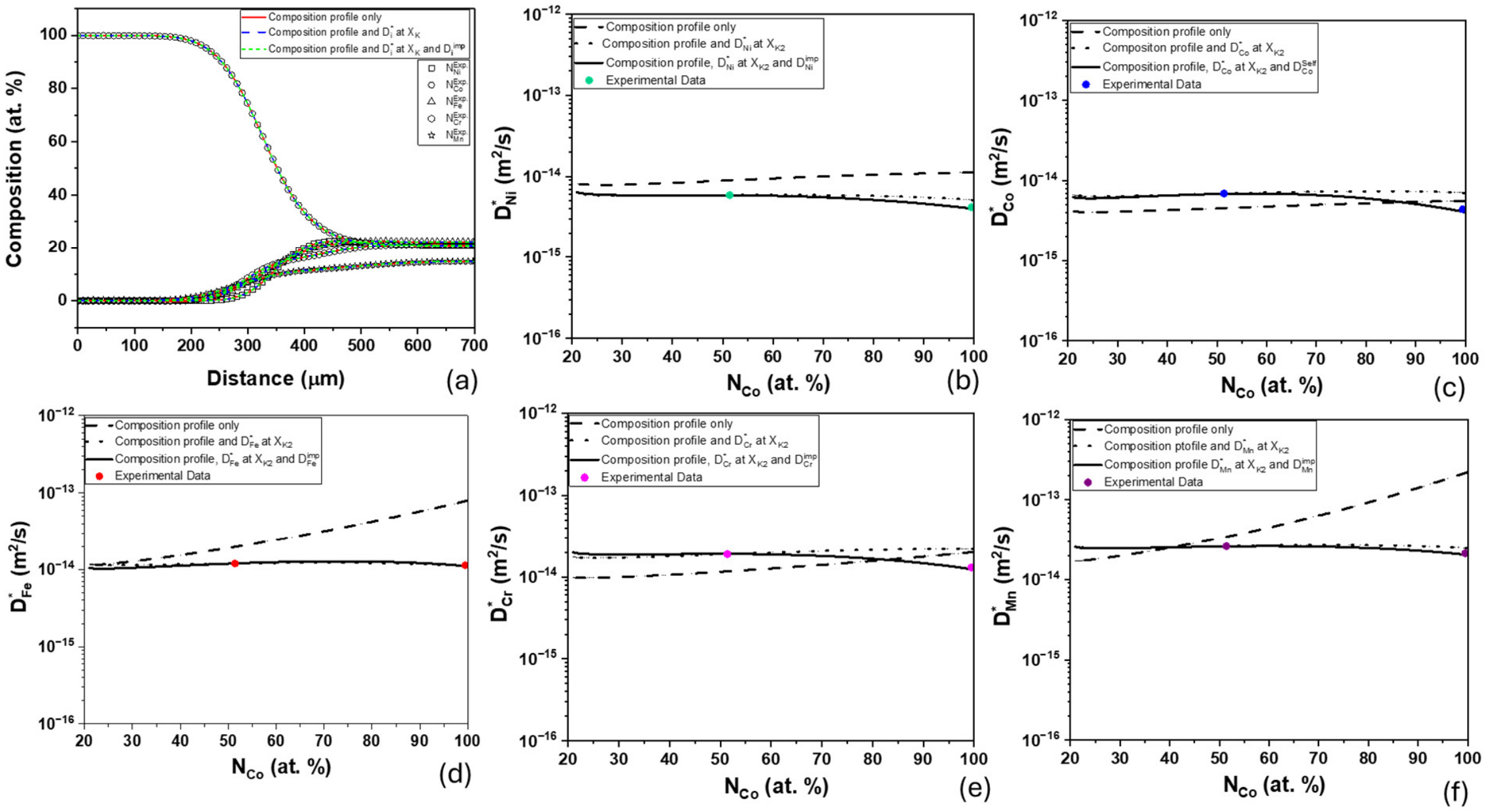

Figure 5 A comparison of PINN optimized (a) diffusion profile of DC1, composition-dependent tracer diffusion coefficients of (b) Ni, (c) Co, (d) Fe, (e) Cr (f) Mn with experimental data.

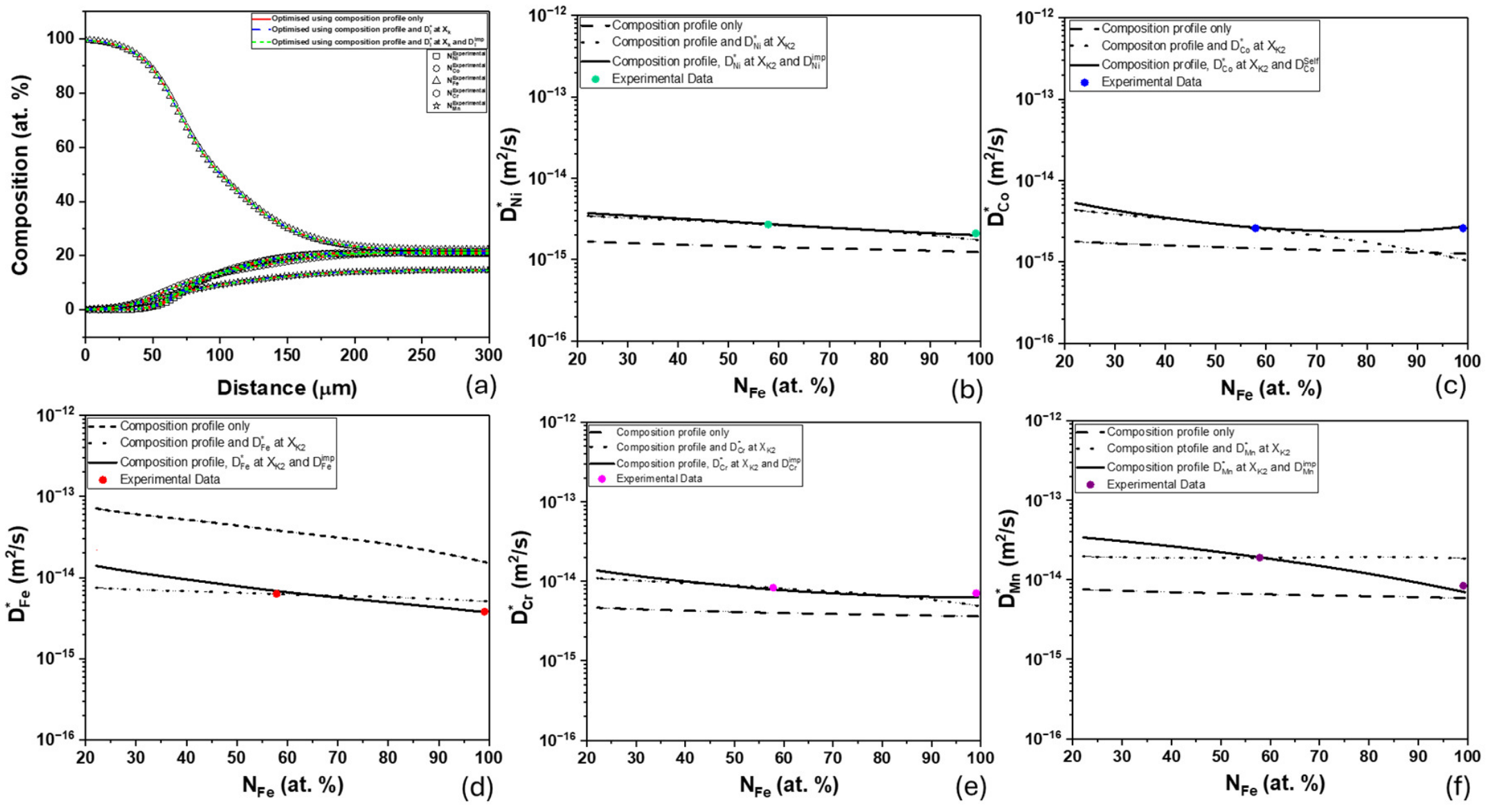

Figure 6 A comparison of PINN optimized (a) diffusion profile of DC1, composition-dependent tracer diffusion coefficients of (b) Ni, (c) Co, (d) Fe, (e) Cr (f) Mn along with impurity diffusion coefficients and the data at the Kirkendall marker plane used as equality constraints.

It is essential to stress that constraint-free optimization can lead to non-physical or misleading mobility databases, even when profile fits can be perfect. Without anchoring the optimization to independently known tracer and impurity diffusivities, the extracted coefficients can reflect mathematically optimal but physically implausible solutions, particularly in the regions where experimental constraints are absent. Therefore, including even sparse experimental diffusivity values significantly shrinks the solution space, guiding the optimization toward a physically consistent solution. The tracer diffusion coefficients estimated at the Kirkendall marker plane and the impurity and self-diffusion coefficients available in the literature are used as the equality constraint. However, there may still be a problem towards the alloy $(NiCoFeCr)_{85}Mn_{15}$ side of the diffusion couple for extracting consistent data, so no diffusion coefficients were used as the equality constraint. To solve this problem, the diffusion couples are produced strategically in this study by coupling the same alloy composition to pure Ni, Co and Fe. Logically, the extracted tracer diffusion coefficients from three diffusion couples towards the alloy $(NiCoFeCr)_{85}Mn_{15}$ should have the same values. However, it may not show this match since diffusion coefficients on the alloy side are not utilized as equality constraints. Therefore, different diffusion couples may show different values, which is indeed found, as shown in Table 7. To solve this problem, we take an average of the values extended from different diffusion couples, as listed in the table, which can be utilized as another equality constraint in the alloy side of the diffusion couples for generating a consistent mobility database. Interestingly, the average values are very similar to the tracer diffusion coefficients estimated by the radiotracer method at the equiatomic composition [35] when extended to the temperature of interest of our study. The difference is in the error range reported by them. The similarity of diffusion coefficients at these two compositions is expected since the composition difference between the end member alloy produced in this study and the equiatomic composition is relatively small. The data in the pure end members as available in the literature [51-59], estimated at the Kirkendall marker planes in this study and at the equiatomic composition by Gaertner et al. [35] or the average diffusivities calculated at the end member alloy (Ni, Co, Fe, $Cr)_{85}Mn_{15}$ are shown on a projected Gibbs triangle, as shown in Fig. 7. For the ease of visualization on a 2D plot, the projected compositions of Ni, Co and Fe ($N_i^p$) on Ni-Co-Fe Gibbs triangle are calculated from actual compositions ($N_i$) by

$$N_i^p(i = Ni, Co, Fe) = \frac{N_i}{1-(N_{Cr}+N_{Mn})} = \frac{N_i}{N_{Ni}+N_{Co}+N_{Fe}} \quad (18)$$

Such that we have $N_{Ni}^p + N_{Co}^p + N_{Fe}^p = 1$.

The visualization of data projected by the Gibbs triangle offers an efficient tool to observe trends in diffusivity across compositional space. The smooth variation seen across the Ni-Co-Fe modified triangle confirms the internal consistency of the dataset and the effectiveness of the PINN-based interpolation. Moreover, this mapping can serve as a starting point for machine learning surrogate models trained on high-quality physical data, opening avenues for fast diffusivity prediction in broader HEA systems.

| | Ni-(NiCoFeCr)$_{85}$Mn$_{15}$ (DC 1) (x $10^{-14}$ m$^2$/s) | Co-(NiCoFeCr)$_{85}$Mn$_{15}$ (DC 2) (x $10^{-14}$ m$^2$/s) | Fe-(NiCoFeCr)$_{85}$Mn$_{15}$ (DC 3) (x $10^{-14}$ m$^2$/s) | (NiCoFeCr)$_{85}$Mn$_{15}$ Average (x $10^{-14}$ m$^2$/s) | Equiatomic composition [35] (x $10^{-14}$ m$^2$/s) |
|---|---|---|---|---|---|
| $D^*_{Ni}$ | 1.1 | 0.64 | 0.45 | 0.73 | 0.95 |
| $D^*_{Co}$ | 1.1 | 0.62 | 0.52 | 0.75 | 1.1 |
| $D^*_{Fe}$ | 4.6 | 1.05 | 1 | 2.21 | 1.4 |
| $D^*_{Cr}$ | 4.8 | 2 | 1.2 | 2.66 | 2.6 |
| $D^*_{Mn}$ | 11 | 2.6 | 3.8 | 5.8 | 3.4 |

Table 7: Calculating average tracer diffusion coefficients at the end member composition by extracting the data from different diffusion couples. A comparison with the data available in literature by extending it to 1200 ºC at the equiatomic composition [35].

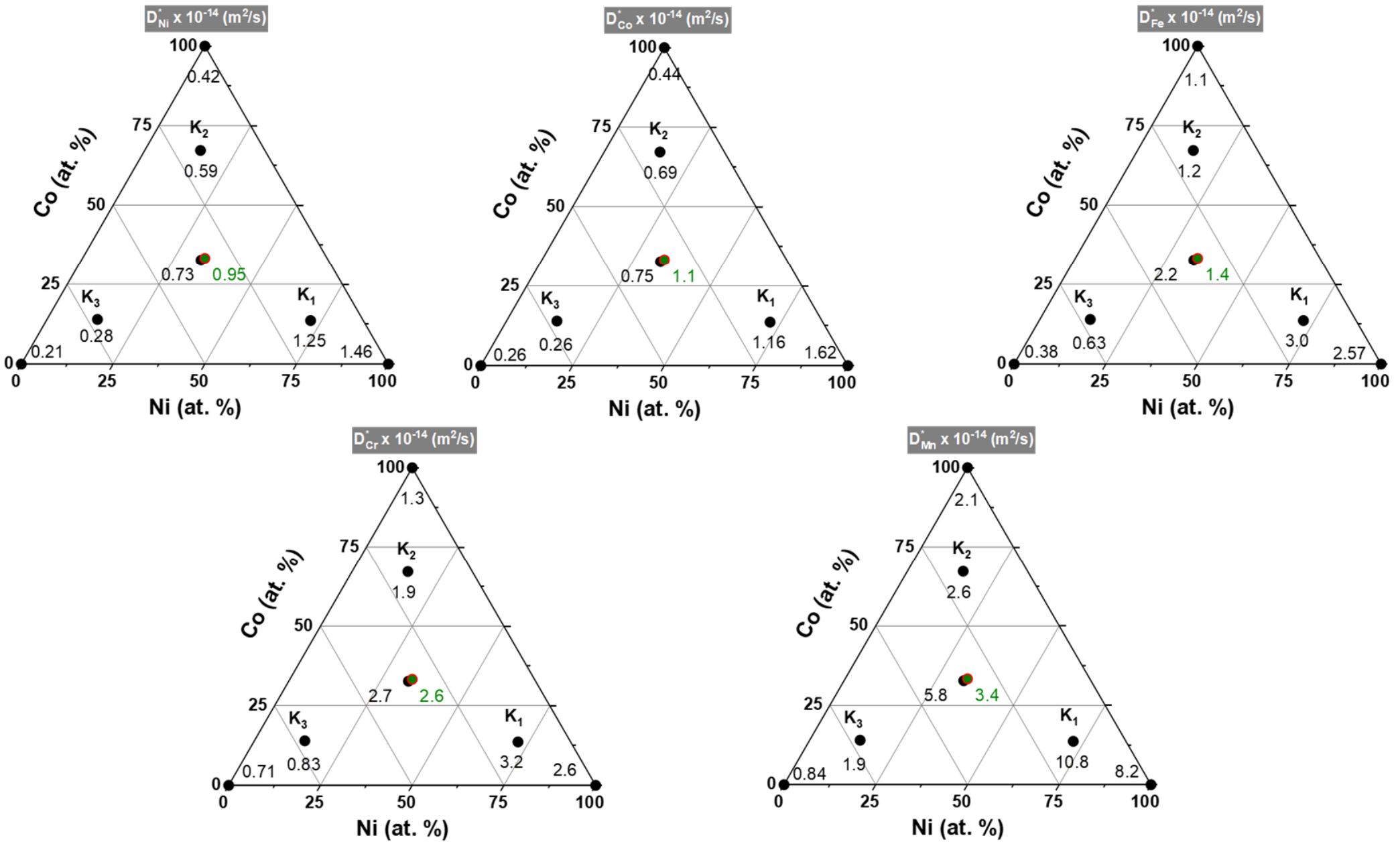

Figure 7: Impurity diffusion coefficients available in the literature [51-59], tracer diffusion coefficients estimated at the Kirkendall marker planes in this study, average tracer diffusion coefficient extracted at the end member composition of (NiCoFeCr)$_{85}$Mn$_{15}$ in this study and at the equiatomic composition after extending the data to 1200 ºC from the literature [35].

As a next step, the average of optimized tracer diffusion coefficients is also utilized as equality constraints along with the self and impurity diffusion coefficients in pure elements, and the data is estimated at the Kirkendall marker plane in this study. It can be seen in Fig. 8 that the optimized composition profile matches the experimentally measured profiles very well, and the tracer diffusion coefficients are now optimized utilizing the tracer diffusion coefficients at two ends and middle of the diffusion couples. The optimized parameters for extracting the tracer diffusion coefficients from three diffusion couples are listed in the supplementary file.

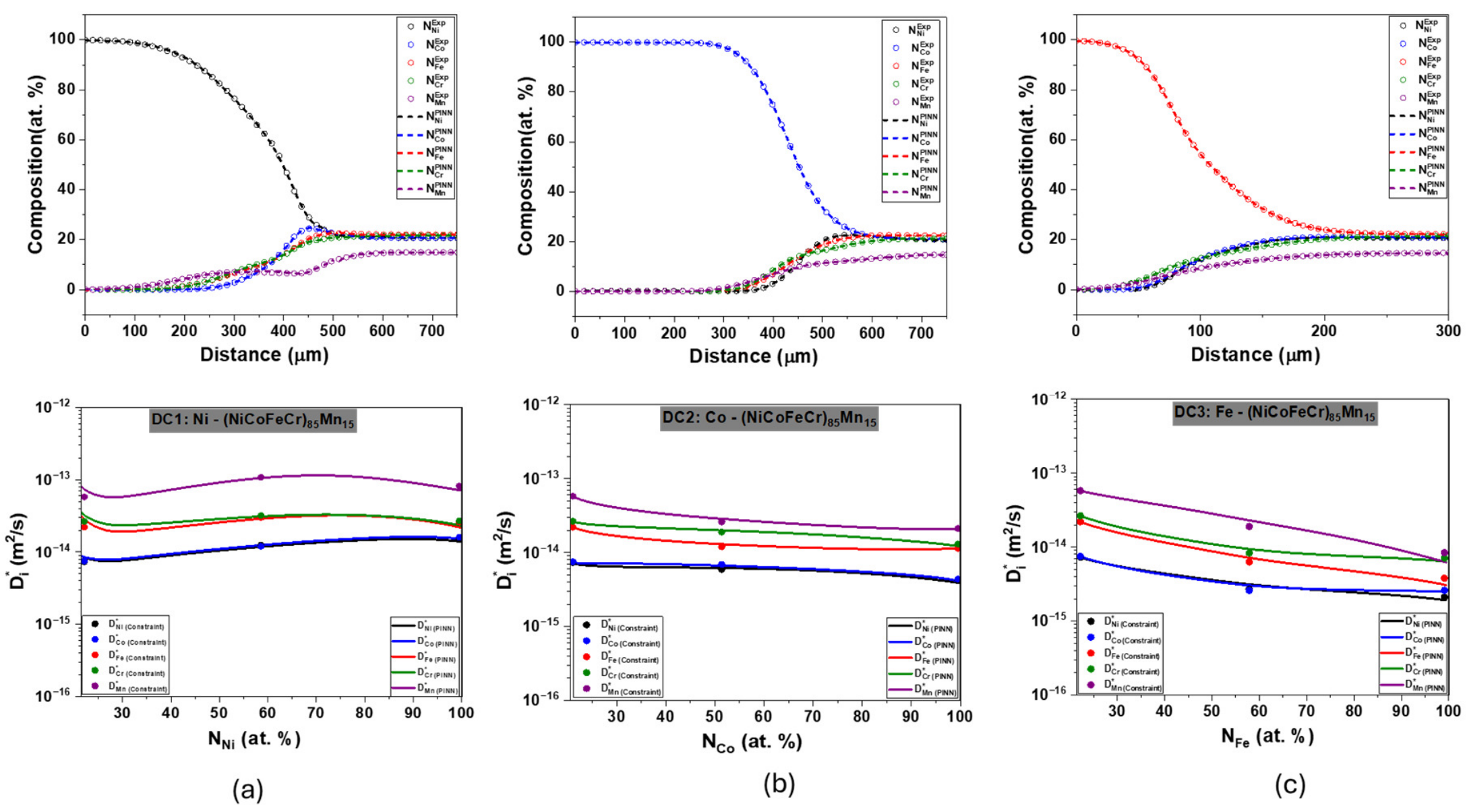

Fig. 8 A comparison of PINN optimized diffusion profile with experimental diffusion profile and optimized tracer diffusion coefficients of elements along with impurity diffusion coefficients, tracer diffusion coefficients at the Kirkendall marker plane and the average diffusion coefficients calculated at the end member composition data used as equality constraints at 1200 ºC for (a) DC1, (b) DC2 and (c) DC3.

We have shown a systematic estimation of diffusion coefficients and optimization for generating a mobility database in a multi-component system. The reliability of experimentally estimated data is very important for the method and analysis described. Mainly, there are two ways to estimate tracer diffusion coefficients:

(i) One can estimate these by intersecting conventional diffusion paths (in which all the elements develop the diffusion profiles) following the Kirkaldy-Lane method [7] or following the augmented Kirkaldy-Lane method by intersecting a conventional diffusion path with a constrained diffusion path or two constrained diffusion paths, as established by Paul's group [12, 13]. The augmented-Kirkaldy method is beneficial in higher-order systems; otherwise, it is not easy to intersect or pass the diffusion paths closely enough, especially in higher-order systems.

(ii) The estimation directly at the Kirkendall, which is demonstrated in this study based on the analysis demonstrated by Paul's group in a multicomponent system [11].

To ensure the quality of diffusion coefficients produced, it is important to compare them with the impurity diffusion coefficients mostly available in the literature or tracer diffusion coefficients measured by the radiotracer method. In this study, we have compared estimated data with both types of data available in the literature on impurity diffusion coefficients [51-59] and tracer diffusion coefficients in the equiatomic composition of NiCoFeCrMn [35] to find similar trends of relative mobilities and systematic variation with composition. One may otherwise report inconsistent data, mainly because estimated interdiffusion coefficients do not clearly indicate the quality of data unless tracer diffusion coefficients with logical values are estimated. For example, the group of van Loo followed the Kirkaldy-Lane method for calculating tracer diffusion coefficients in the Cu-Ni-Fe system [62]. Following, they discussed the issues with the high error in the calculation of interdiffusion coefficients leading to the calculation of random or even illogical values of tracer diffusion coefficients and even with negative values by passing three diffusion profiles from the same composition [63]. They found the illogical values of estimated tracer diffusion coefficients by considering two profiles at a time. This issue could be identified only during the estimation of the tracer diffusion coefficients since interdiffusion coefficients are kind of average, and it is not easy to understand the problems associated with these estimated diffusion coefficients because of faulty diffusion couples. On the other hand, we have shown reliable outcomes following a similar method, i.e. by passing three quaternary diffusion profiles very closely in the NiCoFeCr system [6], by producing three ternary diffusion profiles [11] and a combination of ternary and pseudo-binary diffusion profiles [27] in the Ni-Co-Fe system. Three sets of estimated tracer diffusion coefficients by considering two profiles at a time produced similar tracer diffusion coefficients [6, 11, 27]. The data were comparable to the values that Divinski's group estimated using the radiotracer method [35] in the NiCoFeCr system [6]. We have also shown such reliable outcomes following the augmented Kirkaldy Lane method by producing different types of diffusion paths in the same system [13]. As already explained, the data produced in this study in the NiCoFeCrMn system are also found to be comparable with tracer diffusion coefficients measured by the radiotracer method [35]. When the tracer diffusion coefficients are not available, the comparison of data by producing different types of diffusion couples (the data calculated following the pseudo-binary method or calculated by intersecting different types of diffusion paths) can also instil confidence, which we have done in the Ni-Co-Fe-Cr system [6, 13]. We have noticed the importance of producing reliable diffusion couples compared to unknown errors associated with assessed thermodynamic data in established databanks as the reason behind inconsistent data. This is discussed along with the reliability issues of experimentally estimated data by Nayak and Kulkarni [60] in detail in Ref. [10] since they questioned the quality of the thermodynamic data as a reason for such inconsistency, leading to the comment that the data produced in Ref. [11] by us also should not be considered for mobility database although we have produced reliable and very consistent data. This is led by their first effort with inconsistent data for such calculation in Ref. [60] (as discussed in Ref. [10, 27]) ignoring/disregarding our reliable output in multiple studies validated with the data available in the literature, as already mentioned.

Therefore, in combination with the smart design of diffusion couples by estimating the diffusion coefficients at specific compositions and developing a PINN-based numerical inverse method, we have established an efficient method for mobility data generation over the entire composition range of the diffusion couples. This reduces the experimental burden very significantly since estimating composition-dependent diffusion coefficients over such a composition range, especially in multicomponent space, purely experimentally, is an enormous task to even attempt. Moreover, it extracts tracer, intrinsic and interdiffusion coefficient information from just three well-designed couples, without needing radioactive tracers or numerous intersecting paths. This method is the generic and transferable framework to any HEA and multicomponent systems (even with different crystal structures or thermodynamic databases). Most importantly, it provides a scalable, open-source method, which can establish rapid screening of kinetic trends, train machine-learning surrogate models, inform CALPHAD-type mobility assessments in data-scarce regions, and serve as reliable input for mesoscale simulations such as phase-field modelling of microstructural evolution in HEAs and other multicomponent alloys.

## 4. Conclusion

The outcome of this study based on experimental analysis and PINN optimizations can be summarised as:

(i) A strategic design of diffusion couples is demonstrated in combination with PINN-based optimization to generate a mobility database over the whole composition range of diffusion couples. The NiCoFeMnCr system has limited solubility of Cr and Mn in the FCC solid solution phase. This can be extended to other systems with complete solubility of all the elements by coupling equiatomic compositions with pure elements. One may also follow another design strategy such that a pure element is coupled with the equiatomic composition of other elements. For example, in a system with elements A, B, C,....N, one can couple A with the equiatomic composition of (B, C,....,N) or B with the equiatomic composition of (A, C,....,N). This will have the advantage of the composition of all the elements zero at one of the end members (as explained in this study and with more detail in Ref. [11, 30]. This is more suitable for calculating tracer diffusion coefficients of all the elements with smaller errors than the diffusion couple designed in this study. However, this is unsuitable for PINN optimization since the diffusion coefficients at the end member with equiatomic compositions may not be known. The design strategy described in this study is more suitable since one can calculate the average diffusion coefficients by extending to end member compositions from different diffusion couples. Such a calculation in this study has shown a very good match with the data available in the literature.

(ii) We have shown the importance of the estimation/calculation of tracer diffusion coefficients for generating a reliable mobility database using the numerical inverse method. Most studies in ternary and multicomponent systems published until now reported only the interdiffusion coefficients. As explained, the discussion of diffusional interactions considering the interdiffusion coefficients can be misleading (especially in a concentrated alloy, for example, medium and high entropy alloys). Only the intrinsic diffusion coefficients indicate the true value of the diffusion coefficient of a particular element, indicating the correct diffusional interactions. Interdiffusion coefficients are important to correlate directly with the diffusion profiles but vague in nature since these are kind of average of intrinsic diffusion coefficients.

(iii) Vacancy wind effect is found to be significant in alloys with relatively high concentrations of alloying elements and should not be ignored. This significantly influences certain cross-intrinsic diffusion coefficients.

(iv) For the first time, a robust Physics-Informed Neural Network (PINN) has been successfully developed and demonstrated for extracting composition-dependent tracer diffusivities in a five-component (HEA) system. It embeds Fick's second law via Boltzmann-transformed ODEs, Onsager formalism, and thermodynamic factors in a constrained optimization framework. This Constraint-enforced inverse modelling for *physical realism* of the extracted diffusion coefficients shows that optimizing only against diffusion profiles leads to unreliable or non-unique tracer diffusivities. Including experimental tracer data (at Kirkendall marker plane) and impurity diffusivities as equality constraints are critical for reliable mobility database construction.

(v) High-dimensional optimization (75+ parameters) handled successfully, in which tracer diffusivity modelled as a 15-term polynomial per element across a 5-component system. PINN is trained using adjoint-based automatic differentiation and empirically tuned loss weighting. The Bayesian hyperparameter optimization improves network convergence and generalization. Importantly, it avoids overfitting via regularization and staged training**,** enabling convergence even from random initial guesses without prior knowledge of sub-systems**.**

(vi) The combined experimental estimation of diffusion coefficients from a single diffusion profile and constraint-enhanced PINN numerical method facilitates the extraction of reliable composition-dependent diffusion coefficients over a very large composition range depending on the composition range of the diffusion couples.